\documentclass[aps,preprint,tightenlines,superscriptaddress,
amsfonts,amssymb,amsmath,byrevtex,showpacs,nofootinbib,11pt]{revtex4-1}

\usepackage[polish,english]{babel}
\usepackage[applemac]{inputenc}
\usepackage[T1]{fontenc}

\usepackage{latexsym}
\usepackage{graphicx}
\usepackage{geometry}
\usepackage{epstopdf}
\usepackage{subfig}
\usepackage{color}

\usepackage{xcolor}
\usepackage{pdfsync}
\usepackage{fancyhdr}
\usepackage{makeidx}
\usepackage[breaklinks,colorlinks,backref]{hyperref}
\hypersetup{
    colorlinks, 
    linktoc=all, 
    linkcolor=blue,  
  citecolor=hyptxt,
  urlcolor=blue}
  \hypersetup{
  citebordercolor=,
  filebordercolor=green,
  linkbordercolor=blue
}
\definecolor{hyptxt}{rgb}{0.7, 0.4, 0.9}
\usepackage{bibtopic}

\newcommand{\be}{\begin{equation}}
\newcommand{\ee}{\end{equation}}

\begin{document}

\title{Quantum origin of the Minkowski space}

\author{Grzegorz Plewa\footnote{Email: {\tt greq771@gmail.com}}}
\affiliation{National Centre for Nuclear Research, Pasteura 7, 02-093 Warsaw, Poland}

\begin{abstract}
We show that $D=4$ Minkowski space is an emergent concept related to a class of operators in extended Hilbert
space with no positive-definite scalar product. We start with the idea of position-like and momentum-like operators (Plewa 2019 {\it J. Phys. A: Math. Theor.} 52 375401), introduced discussing a connection between quantum entanglement and geometry predicted by ER=EPR conjecture. We examine eigenequations of the simplest operators and identify $D=4$ Minkowski space as to be spanned by normalized eigenvectors corresponding to the zero eigenvalue. Both spacetime dimension and signature of the metric are fixed by
the regularization procedure. We generalize the result to the case of more general operators, being analogues to quantum
fields. We reproduce the Minkowski space again, however, now in a holographic way, as being identified with the conformal
boundary of $AdS_5$. We observe an interesting analogy to string theory and, in particular, to AdS/CFT correspondence.
\end{abstract}

\maketitle

\tableofcontents

\section{Introduction}

It has been proposed that classical spacetime may not be a fundamental concept, but having a quantum origin.
Earlier observations presented in \cite{VanRaamsdonk:2009ar,VanRaamsdonk:2010pw} suggest that quantum entanglement
could be a quantum notion originating space. More precisely, using Ryu-Takayanagi formula \cite{Ryu:2006bv} it was
shown that connectivity of space is related to quantum entanglement. This suggests a sort of duality between quantum
states and spacetime. A realization of such a duality is the ER=EPR conjecture
\cite{Maldacena:2013xja,Susskind:2014yaa,Susskind:2014moa,Susskind:2016jjb}, stating that there is a link between
connected back hole solutions and maximally entangled quantum states. Itself, the conjecture serves as a potential
resolution of AMPS firewall paradox \cite{Almheiri:2012rt}. From a broader perspective, ER=EPR provides even wider
connection between entangled particles and wormholes. In the strongest form, it predicts that every maximally
entangled state can be interpreted geometrically in terms of some (connected) black holes. Still, quantum
entanglement may be ``not enough'' \cite{Susskind:2014moa}, and additional notions, like quantum complexity \cite{Susskind:2014moa,Brown:2015bva,Brown:2017jil,Susskind:2018pmk,Chapman:2018hou} are required. In particular,
quantum complexity seems to be essential in understanding quantum nature of the growth of the Einstein-Rosen
bridge \cite{Susskind:2018pmk}. One expects a significant modification of a classical black hole geometry for
sufficiently long time scales, caused by quantum processes behind the horizon. All of these facts make black
holes quantum objects, revealing interesting connections between quantum mechanics and gravity.

It was also suggested that gravity and quantum mechanics are not separate subjects \cite{Susskind:2017ney}.
Instead, they are strongly correlated and this holds true even far away from the fundamental scale: gravity
is needed for consistent description of seemingly non-gravitational systems. To some extent, this is already
the case within the AdS/CFT correspondence \cite{Maldacena:1997re,CasalderreySolana:2011us,Hubeny:2014bla}.
Strongly coupled states of conformal sypersymmetric Yang-Mills theory translate into weakly coupled
gravitational solutions in (effectively) five-dimensional spacetime. The main difference is that this
holographic description has nothing to do with ``ordinary'' gravity in asymptotically flat four-dimensional
Minkowski space. It is possible, however, that further generalization of various dualities can eventually
result in more general gravitational language, covering even larger class of relativistic states. It could
happen that finding gravitational imprint of various quantum correlations can make these dualities ``real''.
This may uncover the true quantum nature of gravity.

Following motivations presented above and keeping in mind the ER=EPR hypothesis, we continue the analysis
outlined in \cite{Plewa:2018llx} and examine mathematical formalism based on the idea of operators in the
extended Hilbert space. Here the term {\it extended Hilbert space} refers to a generalization  incorporating
negative norm states, i.e. the Hilbert space with no positive definite scalar product
\cite{PhysRev.123.2183,doi:10.1143/PTP.59.972}. As it was shown in \cite{Plewa:2018llx}, one can construct
a special type of operators in that space, which play a role of the standard position and momentum. They
can be used to reformulate the quantum harmonic oscillator in such a way that there is a link with spacetime.
In particular, it was shown that there is a connection between maximally entangled ground state of
a two-dimensional oscillator and the geometry of $AdS_3$.

Rather than discussing the extension of the duality for a larger class of states, we look closer on the
operators themselves, considering their eigenequations. We show that
solutions to these equations are highly non-trivial and can be interpreted geometrically.

This paper is organized as follows. In section \ref{simple} we discuss the eigenequations of the simplest
operators, the {\it position} $e$ and the {\it momentum} $-i \partial\,$ as in \cite{Plewa:2018llx}. We show
that there is an infinite set of finite-dimensional spaces spanned by normalized eigenvalues. We identify
the space corresponding to either zero position or zero momentum to be special because of four-dimensional
orthonormal basis consisting of one negative and three positive-norm states. The space will be then
interpreted geometrically in terms of four-dimensional Minkowski spacetime. We identify  the Poincare
symmetry group as a subgroup of more general complex Lorentz rotations, being symmetry transformations
in the corresponding extended Hilbert space.
In section \ref{generaleig} we consider more general operators labeled  by continuous labels, $e(x)$ and $-i \partial(x)$.
Constructing eigenvectors, we identify an additional constraint equation to get Hermitian operators.
We interpret the equation geometrically as specifying a class of pseudo-hyperbolic spaces \cite{BADITOIU:2002nq}.
We derive $AdS_5$ again as a special solution to the constraint, identifying Minkowski space with the conformal boundary of anti-de Sitter space.
Subsequently, we generalize the results to the case of the most general operators, finding a surprising connection
with string theory. We conclude in section \ref{discussion}.

\section{The simplest operators} 
\label{simple}

\subsection{Operators in the extended Hilbert space}

We start with basic definitions introduced\footnote{In this paper we adopt some small change of terminology regarding operator names.} in \cite{Plewa:2018llx}. 

\newtheorem{def1}{Definition}
\begin{def1}
\label{defi1}
Let $F_0 \subset {\cal{H}}$ be a subspace of some extended Hilbert space $\cal{H}$, i.e. the Hilbert space with no positive-definite scalar product \cite{PhysRev.123.2183,doi:10.1143/PTP.59.972}. We call $F_0$ the {\it fiducial subspace} and assume that it contains at least one normalizable state. Let $\{ e_i,\partial_i \}_{i=1}^{N}$ stands for a set of operators in $\cal{H}$, defined so that  acting on  any $|\phi_a\rangle \in F_0$, one has
\begin{align}
\label{zera}
\langle \phi_b | e_{i_1}^{n_1}...e_{i_k}^{n_k}  | \phi_a \rangle  = \langle \phi_b | \partial_{i_1}^{n_1}...\partial_{i_k}^{n_k}  | \phi_a \rangle:= 0,
\\[1ex]
\label{komutatory}
[e_i,e_j]:=0, \quad [\partial_i,\partial_j]:=0,
\\[1ex]
\label{diffeo1}
\partial_i \, e_{j_1}...e_{j_n} | \phi_a \rangle := \delta_{i j_1} e_{j_2}...e_{j_n} | \phi_a \rangle +...+ \delta_{i j_n} e_{j_1}...e_{j_{n-1}} | \phi_a \rangle,
\\[1ex]
\label{diffeo2}
e_i \, \partial_{j_1}...\partial_{j_n} | \phi_a \rangle := - \, \delta_{i j_1} \partial_{j_2}...\partial_{j_n} | \phi_a \rangle -...- \, \delta_{i j_n} \partial_{j_1}...\partial_{j_{n-1}} | \phi_a \rangle,
\\[1ex]
\label{eherm}
e_i^\dagger:=e_i,
\end{align}
for $N$, $i_k$,  $j_k$  $n_k \in \mathbb{N}$. We call $e_i$ and $\partial_j$ $e$-operators.
\end{def1}
We defined $e$-operators postulating how they act on states belonging to the fiducial subspace $F_0$. As in \cite{Plewa:2018llx}, we restrict to a single normalizable vector in that space, the so-called {\it fiducial vector} $|\phi\rangle$ (still the space $F_0$ itself can be highly non-trivial). Acting with products of either $e_i$ or $\partial_i$ operators results in zero-norm states in the form $e_1|\phi\rangle$, $e_1 e_2|\phi\rangle$, $\partial_2^2 |\phi\rangle$, etc. Despite they are zero norm states, their linear combinations can be normalizable. In particular, a single pair $(e_1,\partial_1)$ of $e$-operators and a single fiducial vector give rise infinitely-dimensional extended Hilbert space. The latter is spanned by positive, negative and zero norm states resulting from products of operators acting on $|\phi\rangle$. Eqs. \eqref{diffeo1}-\eqref{diffeo2} ensure that any product of operators can be expressed solely in terms of either $e_i$ or $\partial_j$. Notice that according to eq. \eqref{diffeo1}, $\partial_i$ looks like ''derivative'' with respect to $e_i$. Similarly, ignoring the minus sign in eq. \eqref{diffeo2}, $e_i$ resembles ''derivative'' with respect to $\partial_j$. This is why we reserved a single name\footnote{In \cite{Plewa:2018llx} they are called {\it external operators}.} {\it $e$-operators} to both $e_i$ and $\partial_j$. 

Following the notations adopted in \cite{Plewa:2018llx} we introduce two spaces, $E[F_0]$ and $D[F_0]$, defined as being spanned by combinations of respectively $e_i$ and $\partial_i$ acting on the fiducial vector but not the fiducial vector itself. For instance, $e_1 e_2|\phi\rangle, e_i^2|\phi\rangle,... \in E[F_0]$, $\partial_1 \partial_2|\phi\rangle, \partial_i^2|\phi\rangle,... \in D[F_0]$, but $|\phi\rangle \notin E[F_0]$ and $|\phi\rangle \notin D[F_0]$. We call the (full) extended Hilbert space\footnote{More specifically, since the definition bases on the choice of the fiducial subspace, one should write ${\cal{H}}[F_0] = F_0 \oplus E[F_0] \oplus D[F_0]$. However, restricting to a single fiducial vector, we intentionally ignore such details.} ${\cal{H}} = F_0 \oplus E[F_0] \oplus D[F_0] $ the {\it maximally extended space}, reserving the term {\it extended space} to $E[F_0] \oplus D[F_0]$. The extended space plays a special role constructing the quantum harmonic oscillator \cite{Plewa:2018llx}.

The important thing is that eq. \eqref{eherm} does not automatically imply that $e_i$ are Hermitian in the standard sense, i.e.
\be
\label{stand_herm}
(u, e_i v) = (e_i u,v),
\ee
for $u,v \in \cal{H}$. Instead, eq. \eqref{eherm} defines a Hermitian conjugate of the object $e_i$. Perhaps it would be better to rewrite eq. \eqref{eherm} as $e^*_i = e_i$. However, because both $e_i$ and $\partial_j$ are operators, following the notation adopted in \cite{Plewa:2018llx}, we will continue to use the standard symbol of Hermitian conjugate. As we shall see, the operators are indeed Hermitian in the corresponding space of eigenstates. Finding that space will be the main subject of this paper.

It turns out that the rule \eqref{eherm} has to be supplemented by the analogous condition for $\partial_i$
\be
\label{dherm}
\partial_i^\dagger:=-\partial_i.
\ee
Rather than being postulated, this is required by consistency of the algebra generated by commutators of $e$-operators (see \cite{Plewa:2018llx} for more details). The scalar product in the maximally extended space is trivially induced by the product in $F_0$, and the rules \eqref{zera}-\eqref{eherm} and \eqref{dherm}. For instance, for two vectors $u = (e_1+\partial_1)|\phi\rangle$ and $v = (e_1-\partial_1)|\phi\rangle$, one finds
\be
(u,v) = \langle \phi| (e_1+\partial_1)^\dagger  (e_1-\partial_1)|\phi \rangle = \langle \phi| e_i^2 |\phi \rangle - \langle \phi| e_1 \partial_1 |\phi \rangle - \langle \phi| \partial_1 e_1 |\phi \rangle + \langle \phi| \partial_i^2 |\phi \rangle = 0.
\ee
Making use of eqs. \eqref{komutatory}-\eqref{eherm} and \eqref{dherm}, leads to the following commutation relation
\be
\label{gendekom}
[\partial_i, e_j] = \delta_{ij} \hat{\eta},
\ee 
where 
\begin{align}
\nonumber
\hat{\eta} | u \rangle &:= 2 | u \rangle \,:  \quad  |u \rangle  \in F_0,
\\[1ex]
\hat{\eta} | u \rangle &:= | u \rangle  \,: \quad  |u \rangle  \in E[F_0] \oplus D[F_0].
\end{align}
Notice that $\hat{\eta} = id$ in the extended space $E[F_0] \oplus D[F_0]$. Eq. \eqref{gendekom} simplifies in that space becoming $[\partial_i, e_j] = \delta_{ij}$. The last allows to interpret $e_i$ as analogue of the position operator, while $-i \partial_i$ plays a role of the momentum. From now on we will call them respectively the $e$-position and $e$-momentum. 

\subsection{Position eigenequation}
\label{re}

Below we discuss the position eigenequation. Restrict for simplicity to a single pair of operators $(e, \partial)$; here we skipped the labels for convenience. Acting on a given normalized fiducial vector $|\phi \rangle$ the operators span infinitely-dimensional maximally extended space $\cal{H}$. The eigenequation reads
\be
\label{ew}
e |e_\lambda \rangle = \lambda |e_\lambda \rangle.
\ee
Looking for a solution, one may start with the most general form of the vector in the corresponding maximally extended space:
\be
\label{ev0}
 |e_\lambda \rangle = \left( a + \sum_{n=1}^\infty b_n e^n + \sum_{n=1}^\infty c_n \partial^n \right) |\phi \rangle,
\ee
where $a$, $b_n$, $c_n$ are unknown coefficients to be fixed. Inserting the ansatz \eqref{ev0} into the eigenequation \eqref{ew}, and solving the corresponding recurrence equations, gives
\be
\label{eei}
|e_\lambda\rangle = a E_\lambda | \phi \rangle,
\ee
where
\begin{equation}
\label{Eop}
E_\lambda = 1+\sum_{n=1}^{\infty} \left( \lambda^{-n} e^n + \frac{(-\lambda)^n}{n!} \partial^n \right),
\end{equation}
and $a$ stands for the overall normalization coefficients. The recurrence equations fix both $b_n$ and $c_n$, leaving $a$ as a free parameter. Being irrelevant from perspective of the eigenequation, the coefficient $a$ is expected to be fixed by the normalization condition (if the eigenvector is normalizable). We ignore this for a moment, letting temporarily $a=1$. The product of two eigenvectors \eqref{eei}, labeled by two eigenvalues $\lambda_i$, $\lambda_j$, reads
\begin{equation}
\label{Eij}
\langle e_{\lambda_i} | e_{\lambda_j} \rangle = 1 + \sum_{n=1}^{\infty} \left[ \left( \frac{\lambda^*_i}{\lambda_j}  \right)^n + \left( \frac{\lambda_j}{\lambda^*_i}  \right)^n  \right],
\end{equation}
where ''$*$'' stands for complex conjugate. Clearly, $\lambda_i$ should be real for Hermitian operator. We kept the complex conjugate in eq. \eqref{Eij} because we have not proved that $e$ is Hermitian. If so, eigenvectors from different eigenvalues $\lambda_i \neq \lambda_j$ should be orthogonal.

Contrary to what is expected, the product \eqref{Eij} implies that this is not the case unless $|\lambda_i| = |\lambda_j|$. If $|\lambda_i| \neq |\lambda_j|$, the product is divergent and the eigenvectors are not orthogonal. What is more, they are non-normalizable, i.e. $\| e_\lambda \| = \infty$. Despite the latter is common for operators with continuous spectrum (they are mostly non-normalizable), the former means that $e$ is not Hermitian in the whole maximally extended space $\cal{H}$. Still, it could happen the operator is Hermitian if it is restricted to some subspace ${\cal{E}} \subset {\cal{H}}$. Keeping this in mind, we call ${\cal{E}}$ the {\it restricted Hilbert space} (within $\cal{H}$). If exists, the space is of special importance because it makes $e$-position a candidate for observable. Since ${\cal{H}}$ is not positive-definite, no one guaranties that ${\cal{E}}$ will be free of negative norm states. In fact, this is in parallel analogy to construction of the harmonic oscillator presented in \cite{Plewa:2018llx}. Hilbert space of the quantum harmonic oscillator emerges as a natural positive-definite sector in the extended space. Finding that sector, one finds the oscillator. Similarly, we now expect that constructing  the restricted Hilbert space ${\cal{E}}$ will get some insight into physical meaning of $e$-position.

Having said that, we are now ready to start the construction. Since, by definition, $e$ is Hermitian in ${\cal{E}}$, the eigenvalues should be real, i.e. $\lambda \in \mathbb{R}$ in eq. \eqref{Eij}. Another observation is that despite the eigenvectors \eqref{eei} are not normalizable, they can be easily regularized. In order to do so, consider a finite cut-off $\Lambda \gg 1$ and the following regularized version of the operator \eqref{Eop}
\begin{equation}
\label{Eop2}
E_\lambda^\Lambda := 1+\sum_{n=1}^{\Lambda} \left( \lambda^{-n} e^n + \frac{(-\lambda)^n}{n!} \partial^n \right).
\end{equation}
Eigenvectors \eqref{eei} can be now rewritten as $|e_\lambda\rangle = \lim_{\Lambda \rightarrow \infty } E_\lambda^\Lambda | \phi \rangle$. It is wort to underline that vectors constructed with the help of the operator \eqref{Eop2} with finite $\Lambda$ are not the exact position eigenstates. In fact, the reason for introducing the form \eqref{Eop2} consists in parameterization of the limit $\Lambda \rightarrow \infty$. Rather than imposing a real cut-off, we will let $\Lambda \rightarrow \infty$ in the end. Note that the operator \eqref{Eop2} is defined up to the overall normalization constant and any vector in the form $a \lim_{\Lambda \rightarrow \infty }E_\lambda^\Lambda | \phi \rangle$, with $a \in \mathbb{C}$, is an eigenvector as well.  One can formally make $a$ infinitesimally small dividing by infinite length and requiring the resulting state will be finite. This can be done introducing ''regularized'' eigenvectors in the form
\begin{equation}
\label{ee2}
|e_\lambda^r\rangle  :=  \lim_{\Lambda \rightarrow \infty} \frac{1}{\sqrt{2 \Lambda}} E_\lambda^\Lambda | \phi \rangle.
\end{equation}
The latter is to be supplemented by a subtle redefinition of the product:
\begin{equation}
\label{prod2}
\langle e_{\lambda_i}^r|e_{\lambda_j}^r \rangle := \lim_{\Lambda \rightarrow \infty} \left( \frac{1}{2 \Lambda} \langle \phi | (E^\Lambda_{\lambda_i})^\dagger E^\Lambda_{\lambda_j} | \phi \rangle \right).
\end{equation}
It is straightforward to observe that eqs. \eqref{ee2}-\eqref{prod2} imply the eigenvectors are unit, i.e.  $\langle e_{\lambda}^r|e_{\lambda}^r \rangle = 1$. Alternatively, one can define the vectors \eqref{ee2} as $|e_\lambda^r\rangle  :=  \lim_{\Lambda \rightarrow \infty} \frac{1}{\sqrt{2 \Lambda_0}} E_\lambda^\Lambda | \phi \rangle$, supplementing the definition with the following version of product 
\be
\langle e_{\lambda_i}^r|e_{\lambda_j}^r \rangle := \lim_{\Lambda \rightarrow \infty} ( \langle e_{\lambda_i}^r|e_{\lambda_j}^r \rangle |_{\Lambda_0 = \Lambda} ).
\ee
For the sake of simplicity we adopt the form \eqref{ee2} of regularized eigenvectors, understanding they are normalized according to the rule  \eqref{prod2}. 

One can check that eigenvectors \eqref{ee2} corresponding to opposite non-zero eigenvalues are orthogonal, i.e. $\langle e_{-\lambda}^r|e_{\lambda}^r \rangle = 0$, $\lambda \neq 0$. However, this is not the case in general. Fixing a single $|\lambda| = \lambda_0 > 0$, one specifies a two-dimensional space ${\cal{E}}_{\lambda_0} \subset {\cal{H}}$ spanned by eigenvectors of two opposite eigenvalues $\lambda = \pm \lambda_0$. Clearly, ${\cal{E}}_{\lambda_0}$ is an ordinary two-dimensional Hilbert space and $e$ is Hermitian in that space. The positive real parameter $\lambda_0$ is arbitrary, but should be fixed. Once this is done, all other eigenvalues $|\lambda| \neq \lambda_0$ are forbidden. Still, none of the values $\lambda_0>0$  is special. One can formally incorporate all of them considering direct sum of the corresponding spaces ${\cal{E}}_{\lambda_0}$
\be
\label{eps}
{\cal{E}}_{+} =  \bigoplus_{\lambda_0 > 0} {\cal{E}}_{\lambda_0},
\ee
and supplementing ${\cal{E}}_{+}$ with an additional superselection rule eliminating superpositions of states from different absolute value of the eigenvalues. In what follows, $|\lambda|$ plays a role of the conserved charge. The only allowed superpositions involve eigenvectors corresponding to opposite eigenvalues. This guaranties that the $e$-position is Hermitian. All superselection sectors in eq. \eqref{eps}, despite separate, are equivalent in the sense there is a trivial isomorphism preserving the scalar product. In fact, all the sectors are (ordinary) two-dimensional Hilbert spaces. 

The construction above may seem to be unfamiliar from quantum mechanical point of view, since constructing the restricted Hilbert space $\cal{E}$ involves additional superselection rules. As we shall see, they have a deeper geometrical meaning.

\subsection{Generalization to complex eigenvalues}
\label{comp}

So far we have restricted ourselves to non-zero eigenvalues. It turns out, however, the case $\lambda=0$ will be interesting from geometrical point of view. Before discussing the limit $\lambda \rightarrow 0$, it will be convenient ignoring for a moment the fact we construct a space where $e$ is manifestly Hermitian. Instead, we temporarily allow general complex eigenvalues. The reason is that we would like to enlarge the space \eqref{eps}, incorporating eigenvectors from arbitrarily small but complex eigenvalues. 

As in the case of eq. \eqref{ee2}, we start considering a pair of normalized eigenvectors. Letting in eq. \eqref{Eij} $\lambda_i = i$, one finds
\be
\label{clim}
\| e_{\lambda_i} \| =  \lim_{\Lambda \rightarrow \infty} (-1)^{\Lambda}.
\ee
Again, this is ill-defined. On the other hand, the right hand side of the equation above can be easily regularized letting $\Lambda$ to be either even or odd. This leads to a pair of normalized eigenvectors, defined as follows
\be
\label{imvec}
|e_i^{(\Lambda_1)} \rangle  :=  \lim_{N \rightarrow \infty} E_i^{\Lambda_1 (N)} | \phi \rangle, \quad |e_i^{(\Lambda_2)} \rangle  := \lim_{N \rightarrow \infty} E_i^{\Lambda_2 (N)} | \phi \rangle,
\ee
where $\Lambda_1 = \Lambda_1(N) = 2 N$, $\Lambda_2 = \Lambda_2(N)=2 N-1$; $N \in \mathbb{N}$. The limit \eqref{clim} is now well-defined. In particular,
\be
\label{cnorms}
\| e_{i}^{(\Lambda_1)} \| = 1, \quad \| e_{i}^{(\Lambda_2)} \| = -1.
\ee
Note that there is one positive and one negative norm eigenvector. Imposing of the cut-off $\Lambda$ is not regularization in the standard sense, since instead of interpreting the divergent term, we formally allowed all the possible outcomes. Each of them is associated with different normalized vector, making this simple procedure free from ambiguities.

Unfortunately, the product $\langle e_i^{(\Lambda_i)}|e_i^{(\Lambda_j)} \rangle $ is not well-defined due to the presence of double limits:
\be
\label{nun}
\langle e_i^{(\Lambda_1)}|e_i^{(\Lambda_2)} \rangle =  \lim_{N_1 \rightarrow \infty} \lim_{N_2 \rightarrow \infty} \langle \phi | (E_i^{\Lambda_1 (N_1)})^\dagger E_i^{\Lambda_2(N_2)} | \phi \rangle.
\ee
They have to be clarified to get the result. This can be done introducing an order operation $::$, defined as follows
\begin{align}
\nonumber
& : \langle \phi | (E_{\lambda_i}^{\Lambda_i})^\dagger E_{\lambda_j}^{\Lambda_j} | \phi \rangle : \, \, \, \equiv  
\\[1ex]
& \equiv  \frac{1}{2} \langle \phi | (E_{\lambda_i}^{\Lambda_i})^\dagger E_{\lambda_j}^{\Lambda_j} | \phi \rangle \Big |_{\Lambda_i \leq \Lambda_j}  + \frac{1}{2} \langle \phi | (E_{\lambda_i}^{\Lambda_i})^\dagger E_{\lambda_j}^{\Lambda_j} | \phi \rangle \Big |_{\Lambda_i \geq \Lambda_j}.
\label{crule}
\end{align}
We can now define 
\begin{align}
\label{cel}
: \langle e_{\lambda_i}^{(\Lambda_i)}|e_{\lambda_j}^{(\Lambda_j)} \rangle : \, =   \lim_{N_i \rightarrow \infty} \lim_{N_j \rightarrow \infty} : \langle \phi | (E_{\lambda_i}^{\Lambda_i (N_i)})^\dagger E_{\lambda_j}^{\Lambda_j (N_j)} | \phi \rangle : 
\end{align}
In other words, taking the limits we impose additional constraints, i.e. $\Lambda_i \leq \Lambda_j$ and $\Lambda_i \geq \Lambda_j$, treating them in a symmetric way. As it is shown in appendix \ref{cal}, the rule \eqref{cel} transforms the double limit into the following single ones:
\begin{align}
\nonumber
: \langle e_{\lambda_i}^{(\Lambda_i)}|e_{\lambda_j}^{(\Lambda_j)} \rangle :  &= \frac{1}{2} \lim_{N_i \rightarrow \infty}  \langle \phi | (E_{\lambda_i}^{\Lambda_i(N_i)})^\dagger E_{\lambda_j}^{\Lambda_i (N_i)} | \phi \rangle + 
\\[1ex]
&+ \frac{1}{2} \lim_{N_j \rightarrow \infty}  \langle \phi | (E_{\lambda_i}^{\Lambda_j (N_j)})^\dagger E_{\lambda_j}^{\Lambda_j (N_j)} | \phi \rangle,
\label{orule}
\end{align}
and  both they are well defined. For vectors \eqref{imvec} the rule \eqref{orule} gives $: \langle e_i^{\Lambda_i}|e_i^{\Lambda_j} \rangle : =0$. In what follows, the two degenerated eigenvectors are orthogonal. Note that the order \eqref{crule} only matters if $\Lambda_i \neq \Lambda_j$. If they are equal, $: \langle e_{\lambda_i}^{(\Lambda_i)}|e_{\lambda_j}^{(\Lambda_j)} \rangle : \, = \langle e_{\lambda_i}^{(\Lambda_i)}|e_{\lambda_j}^{(\Lambda_j)} \rangle |_{\Lambda_i = \Lambda_j}$.

Having specified exemplary eigenvectors \eqref{imvec} corresponding to the single eigenvalue $\lambda = i$, one can ask if the procedure can be generalized to the case of general complex eigenvalues. As it is shown in appendix \ref{cal2}, this is indeed the case for
\be
\label{phis}
\lambda = r e^{i \varphi}, \quad \varphi = \frac{p \pi}{q},
\ee
where $\frac{p}{q} \in (0,1) \cup (1,2)$, $p,q \in \mathbb{N}$. The corresponding regularized eigenvectors read
\be
\label{imv}
| e_{\lambda}^{(l)} \rangle  = \lim_{N \rightarrow \infty} E_{\lambda}^{\Lambda_{q,l} (N)} | \phi \rangle.
\ee
The eigenvectors are labeled by additional discrete labels $l$ in such a way that
\be
\label{names}
\Lambda_{q,l}(N) = q N+l, \quad l \in \{ 0,1,...,q-1 \}.
\ee
In particular, letting $r=1$, $\varphi = \frac{\pi}{2}$ corresponds to $\lambda = i$, $p=1$, $q=2$ and $l \in \{ 0,1\}$. These two values define two different eigenvectors, specified by two cut-offs, $\Lambda_{2,0}(N) = 2 N$, $\Lambda_{2,1}(N) = 2 N+1$. This reproduces the eigenvectors \eqref{imvec}.

The construction additionally requires that eigenvectors corresponding to different complex eigenvalues does not belong to a single space (see appendix \ref{cal2} for more details). This is a direct analogue of the superselection rule for regularized eigenvectors corresponding to real eigenvalues \eqref{ee2}. However, now each space corresponds to a single complex eigenvalue, while different complex $\lambda$ specify different spaces of regularized eigenvectors. Each space contains one positive, one negative and $q-2$ zero norm states (if $q\leq 2$ then there is no zero norm vectors). The resulted eigenvectors are degenerated, whereas the spaces they belong to are equivalent in the sense they differ only by the number of zero norm eigenstates.

\subsection{Zero eigenvalues and $D=4$ Minkowski space}
\label{zeroeig}

We are now ready to discuss the problem of zero eigenvalues. In order to do so, consider $\lambda = \pm \epsilon$ first, where $\epsilon$ stands for an infinitesimal real parameter. Define
\be
\label{epsre}
|e_\pm \rangle := |e_{\pm \epsilon}^r \rangle,
\ee
where $|e_{\pm \epsilon}^r \rangle$ are given by eq. \eqref{ee2} for $\lambda_0 = \epsilon$. That is, there are two eigenvectors of two opposite eigenvalues $\pm \epsilon$. Both eigenvalues go to zero in the limit $\epsilon \rightarrow 0$. To make the product \eqref{prod2} well-defined we let $\epsilon$ to be arbitrary small, but non-zero. In what follows, the limit $\lambda \rightarrow 0$ consists in treating $\epsilon$ as infinitesimal parameter, and letting $\epsilon \rightarrow 0$ in the end. This is in parallel analogy to introducing regularization parameter $\Lambda$. The only difference is that we first eliminate $\Lambda$, letting $\Lambda \rightarrow \infty$ while keeping $\epsilon$ as a fixed parameter, and then eliminate $\epsilon$ letting $\epsilon \rightarrow 0$. Alternatively, one can keep $\epsilon$ as arbitrarily small but fixed all the time, treating the vectors given by eq. \eqref{epsre} as eigenvectors corresponding to approximately zero eigenvalue. The smaller the value $\epsilon$, the better the approximation. We ignore such details now, however. Physically it does not matter whether some quantity is exactly zero or is arbitrarily small. From now on we interpret eq. \eqref{epsre} as defining a pair of degenerated eigenvectors corresponding to the zero eigenvalue. This value can be found in the limit of either small negative $\lambda = -\epsilon$ or small positive $\lambda = \epsilon$ eigenvalues.

In the next step we generalize the procedure to the case of complex eigenvalues. The idea is to reach $\lambda=0$  by small complex $\lambda$ (i.e. $|\lambda| \ll 1$). For instance, let $\lambda = i \epsilon$. Recalling eqs. \eqref{phis}-\eqref{names}, this corresponds to $r=\epsilon$, $\varphi = \pi/2$, leading to two eigenvectors labeled by $l=0$ and $l=1$. Define
\be
\label{her0}
|e\rangle_0 := |e^{(1)}_{i \epsilon}\rangle, \quad  |e\rangle_1 := |e^{(0)}_{i \epsilon}\rangle,
\ee
where $|e^{(l)}_{i \epsilon}\rangle$ are given by eq. \eqref{imv}. These are defined as respectively negative and positive norm eigenvectors, i.e. ${_0}\langle e|e\rangle_0 = -1$, ${_1}\langle e|e\rangle_1 = 1$. For small but fixed $\epsilon$ the operator $e$ is approximately Hermitian on states \eqref{her0} in the sense that the deviation $|(e^k,e \, e^l)-(e \, e^k,e^l)| = 2 i \epsilon \delta_{kl}$ is small of order $\epsilon$ (here $e^k, e^l$ stand for eigenvectors \eqref{her0}). Letting $\epsilon$ arbitrarily close to zero makes the deviation from Hermiticity, as well as the eigenvalue $\lambda$, arbitrarily small. This is similar to the case of eigenvectors from real eigenvalues \eqref{epsre}. Again, we treat $\epsilon$ as arbitrarily close to zero, eliminating it in the limit $\epsilon \rightarrow 0^+$. The vectors \eqref{her0} turns out to be normalized $e$-eigenvectors corresponding to the zero eigenvalue. In contrast to the vectors \eqref{epsre}, one gets the value $\lambda=0$ in the limit at which the imaginary part of the eigenvalue goes to zero.

Despite the vectors \eqref{epsre} and \eqref{her0} were found using different methods, they are orthogonal. This is due to the prefactor $(2 \Lambda)^{-1/2}$ in the definition \eqref{ee2}. Hence, the two-dimensional space corresponding to $\lambda_0 = \epsilon$, i.e. the space of eigenstates corresponding to opposite eigenvalues $\lambda = \pm \epsilon$, can be effectively enlarged supplementing with vectors \eqref{her0}. This results in the total number of four unit eigenvectors:
\be
|e\rangle_0 := |e^0 \rangle, \quad |e\rangle_1 := |e^1 \rangle, \quad |e\rangle_2 := |e_+ \rangle, \quad |e\rangle_3 := |e_- \rangle.
\label{4base}
\ee
They all correspond to the zero eigenvalue. The first two vectors are  the eigenvactors \eqref{her0}, the last two ones are the eigenvectors \eqref{epsre}. 

At this point one may wonder if the basis \eqref{4base} can be further modified incorporating eigenvectors from different complex (infinitesimal) eigenvalues. The important thing to note is that there is a superselection rule forbidding superposition of states corresponding to different eigenvalues, even if they are arbitrarily close to each other (see appendix \ref{cal2} for more details). Choosing the complex eigenvalue fixes the choice, as in the case of eigenvectors corresponding to real eigenvalues (see eq. \eqref{eps} and the corresponding superselection rule). In particular, instead of the vectors \eqref{her0}, one could consider eigenvectors from different infinitesimally small complex eigenvalues. As was shown in appendix \ref{cal2}, different choices correspond to different number of zero norm states, but this do not affect the number of non-zero norm vectors. Since the zero norm states do not contribute to the scalar product, they are meaningless and can be ignored. In what follows, trying to change regularized eigenvectors, one cannot affect the dimension and signature of the space spanned by the basis of degenerated eigenvectors corresponding to the zero eigenvalue. Increasing dimension is forbidden by superselection rules, while the signature remains the same because we get always one negative and three positive norm eigenvectors. 

Let ${\cal{E}}_0$ stands for the restricted Hilbert space spanned by the basis \eqref{4base} (a subspace of $\cal{H}$ in which $e$ is Hermitian). The set of disconnected spaces \eqref{eps} can be then enlarged by ${\cal{E}}_0$:
\be
\label{eps2}
{\cal{E}} =  \bigoplus_{\lambda_0 \geq 0} {\cal{E}}_{\lambda_0}.
\ee
Being restricted to any of the spaces  ${\cal{E}}_{\lambda_0}$, the operator $e$ is manifestly Hermitian. As in the case of eq. \eqref{eps}, writing down eq. \eqref{eps2} we keep in mind that there are additional superselection rules forbidding superpositions of states corresponding to different $|\lambda|$. 

Among various different spaces in the direct sum \eqref{eps2}, the restricted Hilbert space ${\cal{E}}_{0}$ is special because of the biggest number of orthogonal eigenvectors and the fact that they are all degenerated. Now we will take a closer look at that space. Note that any complex superposition of states corresponding to the zero eigenvalue is an eigenvector as well. Clearly, this is only possible if $\lambda=0$. Let 
\be
\label{ev}
|v\rangle = \xi^\mu | e_\mu \rangle
\ee
be a superposition chosen so that $|\xi^\mu| \ll L$ for $L \gg \epsilon$; $\xi^\mu \in \mathbb{C}$. In what follows, $e |v \rangle = {\cal{O}}(\epsilon)$. The smaller $\epsilon$, the better the approximation. In particular, in the limit $\epsilon \rightarrow 0$, one finds $e |v \rangle \rightarrow 0$. Hence the vector \eqref{ev} is a zero eigenvector in the same sense as any of the basis eigenstates \eqref{4base}. Again, $\epsilon$ is identified with additional regularization parameter, similarly as $\Lambda$. Notice that  $|e\rangle_0$ is a negative norm vector, while the rest ones, $|e\rangle_i$, $i=1,2,3$, have positive norms. This suggest a connection with Minkowski space. Indeed, the basis \eqref{4base} is orthonormal in the sense of the scalar product in Minkowski space, $\langle e_\mu | e_\nu \rangle = \langle e_\nu | e_\mu \rangle = \eta_{\mu \nu}$. Here $\mu,\nu \in \{ 0...3\}$ and  $\eta_{\mu \nu}$ stands for the standard Minkowski metric. However, the scalar product is symmetric under complex Lorentz rotations, i.e. the symmetry group $O(3,1, \mathbb{C})$. Since the negative norm vector is present, we cannot impose the standard normalization condition for the superposition \eqref{ev}. For the same reason in the extended Hilbert space complex amplitudes $\xi^\mu$ have no probabilistic interpretation. Still, as it was shown in \cite{Plewa:2018llx}, in special circumstances they could play a role analogous to the standard probabilities. We ignore this details now, considering the most general symmetry group $O(3,1, \mathbb{C})$.

It turns out that coefficients $\xi^\mu$ can be identified with coordinates in the Minkowski space. To see this note that the symmetry under $O(3,1, \mathbb{C})$ rotations means that
\be
\label{cmin0}
\eta_{\alpha \beta} = \eta_{\gamma \delta} (\Lambda^\gamma_{\, \, \alpha})^* \Lambda^\delta_{\, \, \beta},
\ee
where $*$ stands for complex conjugate. Consider an infinitesimal transformation $\delta \Lambda^\mu_{\, \, \nu} = \delta^\mu_{\, \, \nu}+ \Omega^{\mu}_{\,\, \nu}$, where $\Omega_{\mu \nu} = - \Omega_{\mu \nu}^*$. Extracting real and imaginary parts,
\be
\Omega_{\mu \nu} = \omega_{\mu \nu} + i \, \tilde{\omega}_{\mu \nu},
\ee 
one rewrites the rotation as
\be
\label{loren}
\delta \Lambda^\mu_{\, \, \nu} = \delta^\mu_{\, \, \nu}+ \omega^{\mu}_{\,\, \nu} + i \, \tilde{\omega}^{\mu}_{\,\, \nu},
\ee
where $\omega_{\mu \nu}$, $\tilde{\omega}_{\mu \nu}$ are antisymmetric tensors. The transformation involves twelve real parameters, six for each of two antisymmetric tensors: $\omega_{\mu \nu}$ and $\tilde{\omega}_{\mu \nu}$. Acting on a complex vector $\xi^\mu = \sigma^\mu + i \theta^\mu$; $\sigma^\mu, \theta^\mu \in \mathbb{R}$, the transformation \eqref{loren} gives
\be
\label{tr0}
(\xi^\mu)' = \delta \Lambda^\mu_{\, \, \nu} \xi^\nu =  (\delta^\mu_{\,\, \nu}+ \omega^\mu_{\, \, \nu}) \sigma^\nu - \tilde{\omega}^\mu_{\, \, \nu} \theta^\nu + i \left[ (\delta^\mu_{\,\, \nu}+ \omega^\mu_{\, \, \nu}) \theta^\nu + \tilde{\omega}^\mu_{\, \, \nu} \sigma^\nu \right].
\ee
The real $ (\sigma^\mu)'$ and imaginary $(\theta^\mu)'$ parts of $(\xi^\mu)'$ transform as follows
\begin{align}
 \label{tr1}
(\sigma^\mu)' &= (\delta^\mu_{\,\, \nu}+ \omega^\mu_{\, \, \nu}) \sigma^\nu - \tilde{\omega}^\mu_{\, \, \nu} \theta^\nu,
\\[1ex]
\label{tr2}
(\theta^\mu)' &= (\delta^\mu_{\,\, \nu}+ \omega^\mu_{\, \, \nu}) \theta^\nu + \tilde{\omega}^\mu_{\, \, \nu} \sigma^\nu.
\end{align}
For a fixed $\theta^\mu = \theta^\mu_0=const$, the real part of eq. \eqref{tr0}, given by eq. \eqref{tr1}, is a Poincare transformation
\be
\label{poin1}
(\sigma^\mu)' = (\delta^\mu_{\,\, \nu}+ \omega^\mu_{\, \, \nu}) \sigma^\nu + a^\mu.
\ee
Here $a^\mu = - \tilde{\omega}^\mu_{\, \, \nu} \theta^\nu_0$ plays a role of the translation, while  $\delta^\mu_{\,\, \nu}+ \omega^\mu_{\, \, \nu}$ is a Lorentz rotation. Similarly, for a fixed $\sigma^\mu = \sigma^\mu_0=const$, the imaginary part \eqref{tr2} becomes
\be
\label{poin2}
(\theta^\mu)' = (\delta^\mu_{\,\, \nu}+ \omega^\mu_{\, \, \nu}) \theta^\nu + b^\mu,
\ee
where $b^\mu = \tilde{\omega}^\mu_{\, \, \nu} \sigma^\nu_0$. Again, this is the Poincare transformation  with the same Lorentz rotation and the translation $b^\mu$. In what follows, imposing one of the two following constraints
\begin{align}
\label{const1}
\theta^\mu &= \theta^\mu_0,
\\[1ex]
\label{const2}
\sigma^\mu &= \sigma^\mu_0,
\end{align}
i.e. letting $\theta^\mu=const$ or $\sigma^\mu=const$, respectively the real or imaginary part of the complex vector $\xi^\mu$ transforms under standard four-dimensional Poincare symmetry group. This means that either $\sigma^\mu$ or $\theta^\mu$ can be identified with coordinates in four-dimensional Minkowski space. 

Notice that the constrains \eqref{const1}-\eqref{const2} are essential for this identification. The last fact has a natural interpretation in terms of a higher-dimensional space. Suppose we interpret imaginary parts of the vector $\xi^\mu$ as coordinates in extra dimensions. To this end, consider a vector $\vec{f} = y^a \vec{f}_a$, $a=1,...,8$, where $(y^a):=\{ \sigma^0,...,\sigma^3, \theta^0,...,\theta^3 \}$ are coordinates in an orthonormal, eight-dimensional basis $ \vec{f}_a$. The latter is constructed in such a way that vector spaces spanned by the two subsets $\{ \vec{f}_0,..., \vec{f}_3 \}$, $\{ \vec{f}_4,..., \vec{f}_7 \}$ are isomorphic (in the sense of preserving the scalar product) to the space spanned by the eigenvectors $\{ |e\rangle_0,..., |e\rangle_3 \}$. In what follows, $\vec{f} \in \mathbb{R}^{6,2}$. Imposing one of the two constraints \eqref{const1}-\eqref{const2} results in a four-dimensional hypersurface, the four-dimensional Minkowski space. 

The interpretation above may seem to be artificial because of introduced {\it ad hoc} the eight-dimensional basis $\vec{f}_a$. Still, the construction provides a clear interpretation of the constraints \eqref{const1}-\eqref{const2}. As we shall see discussing more general $e$-operators, the Minkowski space emerges in a similar fashion, being solution to the constraint equation in a higher-dimensional space.

We close this section recalling the form of space of eigenvectors \eqref{eps2}. The fact it splits into separate subspaces is crucial for geometric interpretation of ${\cal{E}}_0$. In particular, this prohibits enlarging the space by additional eigenvectors corresponding to different infinitesimal eigenvalues. The procedure fixes the dimension and signature of the restricted Hilbert space ${\cal{E}}_0$, interpreting it geometrically in terms of  Minkowski spacetime. This means that regularized eigenvectors corresponding to the zero eigenvalue are more like geometric objects, giving rise the Poincare symmetry in four dimensions.

\section{General operators}
\label{generaleig}

Below we  discuss eigenequation of $e$-position labeled by continuous variables, written symbolically as $x = \{ x^0,...,x^{D-1} \}$. In general, generalized $e$-operators $e_i(x)$, $\partial_j(x)$ are given by the following generalization of definition \ref{defi1}:  
\begin{align}
\label{init}
 [e_i(x),e_j(y)]=[\partial_i(x),\partial_j(y)]:=0,
\\[1ex]
 \langle \varphi | e_{i_1}^{n_1}(x_1)...e_{i_k}^{n_k}(x_k)  | \phi \rangle = \langle \varphi | \partial_{i_1}^{n_1}(x_1)...\partial_{i_k}^{n_k}(x_k)  | \phi \rangle := 0,
\\[1ex]
 e_i^\dagger(x) := e_i(x), \quad \partial_i^\dagger(x) := -\partial_i(x),
\\[1ex]
 \partial_i(x) e_j(y) | \phi \rangle := \delta_{ij}\delta^D(x-y)| \phi \rangle.
\end{align}
Additionally we require
\begin{align}
\nonumber
\partial_i(x) e_{j_1}(y_1)...e_{j_n}(y_n)|\phi \rangle &= \delta_{i  j_1} \delta^D(x-y_1) e_{j_2}(y_2)...e_{j_n}(y_n)|\phi \rangle +...
\\[1ex]
\label{gchain1}
&+ \delta_{i j_n} \delta^D(x-y_n) e_{j_1}(y_1)...e_{j_{n-1}}(y_{n-1})|\phi \rangle,
\\[1ex]
\nonumber
e_i(x) \partial_{j_1}(y_1)...\partial_{j_n}(y_n)|\phi \rangle &= -\delta_{i j_1} \delta^D(x-y_1) \partial_{j_2}(y_2)...\partial_{j_n}(y_n)|\phi \rangle +...
\\[1ex] 
\label{gchain2}
&-\delta_{i j_n}\delta^D(x-y_n) \partial_{j_1}(y_1)...\partial_{j_{n-1}}(y_{n-1})|\phi \rangle.
\end{align}
It can be verified that operators satisfy the following commutation relation
\begin{equation}
\label{genxpalgebra2}
[\partial_j(y),e_i(x) ] = \delta_{ij} \delta^D(x-y)\hat{\eta}.
\end{equation}
Linear combinations of $e_i(x)$ and $\partial_j(y)$ leads to creation $\alpha_i^\dagger(x)$ and annihilation $\alpha_i(x)$ operators, satisfying
\begin{align}
\label{alga}
&[\alpha_i(x),\alpha_j^\dagger(y)] = \delta_{i j} \delta^D (x-y) \hat{\eta},
\\[1ex]
\label{alg2b}
&[\alpha_i(x),\alpha_j(y)] = [\alpha_i^\dagger(x),\alpha_j^\dagger(y)]=0.
\end{align}
In the extended space, i.e. the space spanned by combinations of $e$-operators acting on the fiducial vector but not the ''bare'' fiducial vector itself, eq. \eqref{alga} simplifies taking the form of the standard algebra for creation and annihilation operators, i.e.
\be
\label{algebra}
[\alpha_i(x),\alpha_j^\dagger(y)] = \delta_{i j} \delta^D (x-y),
\ee
whereas the commutation relations \eqref{alg2b} remain unchanged. Consider for simplicity operators labeled only by continuous labels, i.e. $e(x)$, $\partial(x)$. For instance, one can identify identify $e(x) = e_1(x)$, $\partial(x) = \partial_1(x)$, rewriting eq. \eqref{algebra} as
\be
\label{algebra2}
[\alpha(x),\alpha^\dagger(y)] = \delta^D (x-y).
\ee
The most general $e$-operators will be discussed in the end of this paper. The eigenequation of $e(x)$ operator reads
\be
\label{eigx}
e(x) | e(x)\rangle = \lambda(x) | e(x)\rangle.
\ee
Here we cannot identify $\lambda(x)=x$, because for $D>1$ this would make no sense from perspective of the eigenequation ($\lambda$ is a scalar-like object). Instead, we have to assume that eigenvalues are given by a scalar function $\lambda(x) = \lambda(x^0,...,x^{D-1})$. We do not fix the dimension $D$ now, letting it to be a free parameter.

The eigenvector $|e(x)\rangle$ is a direct analogue of the standard position eigenstate in position representation $|x\rangle$, i.e.
\be
\hat{x}|x\rangle = x |x\rangle.
\ee
Here
\be
\label{dxpos}
\langle x|x'\rangle  = \delta^3(x-x').
\ee
Being non-normalizable, eigenvectors $|x\rangle$ are orthogonal. Keeping this in mind, we now examine the position $e(x)$. The solution of eq. \eqref{eigx} takes the form 
\be
\label{eex}
| e(x)\rangle = E_\lambda(x) |\phi\rangle,
\ee
where $|\phi\rangle$ is a fiducial vector and $E_\lambda(x)$ stands for the most general combination of (generalized) $e$-operators:
\begin{align}
\nonumber
E_\lambda(x) = a(x) + \sum_{n=1}^\infty \int d^D x_1 ... d^D x_n \, b_n(x;x_1,...,x_n) e(x_1) ...  e(x_n) +
\\[1ex]
+\sum_{n=1}^\infty \int d^D x_1 ... d^D x_n \, c_n(x;x_1,...,x_n) \partial(x_1)  ...  \partial(x_n).
\label{anes}
\end{align}
Here $a(x)$, $b_n(x;x_1,...,x_n)$, $c_n(x;x_1,...,x_n)$ are functional coefficients to be fixed. Supplementing eq. \eqref{eex} with the ansatz \eqref{anes}, and then substituting the resulting vector $| e(x)\rangle$ into the eigenequation \eqref{eigx}, leads to recurrence equations for the functional coefficients. Solving them, one finds
\begin{align}
\label{bcon}
b_n(x;x_1,...,x_n) &= \frac{a(x)}{\lambda^n(x)} \prod_{i=1}^n \delta^D(x-x_i),
\\[1ex]
\label{ccon}
c_n(x;x_1,...,x_n) &= a(x) \frac{(-\lambda(x))^n}{n!},
\end{align}
Here $\lambda(x)$, $a(x)$ are remaining free functional coefficients. Consider the product $\langle e(x')| e(x) \rangle = \langle \phi| E_\lambda(x')^\dagger E_\lambda(x) |\phi \rangle$. Making use of eqs. \eqref{anes}-\eqref{ccon} gives
\be
\label{eprod}
\langle e(x')| e(x) \rangle = a^*(x') a(x) \left( 1+ \sum_{n=1}^\infty \left[  \left( \frac{\lambda(x)}{\lambda^*(x')}\right)^n + \left( \frac{\lambda^*(x')}{\lambda(x)}\right)^n \right]  \right).
\ee
Notice that letting $a(x)=1$, $\lambda(x) = \lambda_j$ and  $\lambda(x') = \lambda_i$, we restore the form \eqref{Eij}. This is expected since with the lack of manifest dependence on continuous variables $x,x'$, the product should reduce to what has been found for $e$ operator. On the other hand, there are two remaining functional parameters in eq. \eqref{eprod}: the eigenvalue $\lambda(x)$ and $a(x)$. 

As in the case of $e$ operator, we will now search for a restricted Hilbert space in which $e(x)$ is Hermitian. In order to do so, we assume that $\lambda(x)$ is real and non-zero. We will not reconsider the problem of zero eigenvalues. In fact, due to relative similarity between eq. \eqref{eprod} and eq. \eqref{Eij}, we would  get nothing interesting. Instead, we will concentrate on a small region in the neighborhood of two close points $x^A$ and $x^A+d x^A$, where $x^A$ stand for continuous labels of $e$-operators ($A \in \{ 0,...,D-1\}$). We ask what guaranties that $e(x)$ is Hermitian in that region. 

Assume that $a(x)=const$, letting $a(x)=1$ in eq. \eqref{eprod}, and impose a finite cut-off $\Lambda$ for $E_\lambda(x)$ in eq. \eqref{anes}. The cut-off will be eliminated by taking the limit $\Lambda \rightarrow \infty$ in the end. The regularized product reads
\be
\label{eprod2}
\langle e_\lambda(x') | e_\lambda(x) \rangle = \left( 1+ \sum_{n=1}^\Lambda \left[  \left( \frac{\lambda(x)}{\lambda(x')}\right)^n + \left( \frac{\lambda(x')}{\lambda(x)}\right)^n \right]  \right),
\ee
where $(x')^A = x^A + d x^A$. We will treat the arguments $x, x' \in \mathbb{R}^D$ as dimensionless parameters. Performing the sum \eqref{eprod2} explicitly and expanding the result in $d x^A$  gives
\be
\label{eprod2b}
\langle e_\lambda(x') | e_\lambda(x) \rangle =  2 \Lambda \left( 1+ \left( \frac{ \Lambda^2}{6} + \frac{\Lambda}{4}+\frac{1}{12} \right) \frac{\partial_{A B} \lambda(x)}{\lambda(x)} d x^A d x^B  \right) +{\cal{O}}(dx^3).
\ee
If $\Lambda \gg 1$, one can neglect two terms in the internal bracket, rewriting eq. \eqref{eprod2} in a bit simpler form
\be
\label{eprod3}
\langle e_\lambda(x') | e_\lambda(x) \rangle =  2 \Lambda \left( 1+ \Lambda^2 \Theta_{A B}(x) d x^A d x^B  \right),
\ee
where
\be
\label{thetaform}
 \Theta_{\mu \nu}(x) := \frac{\partial_{\mu \nu} \lambda(x)}{6 \lambda(x)}.
\ee
Here we skipped the terms of order ${\cal{O}}(dx^3)$\footnote{More specifically, these terms are of order $ {\cal{O}}(\Lambda^2 dx^3)$ and are small if $\Lambda^2 dx^3$ is also small. In particular, this would be the case for $|dx| = {\cal{O}}(\Lambda^{-1})$. The regularization parameter $\Lambda$ is then interpreted as inverse of $|dx|$. Once $|dx| \ll 1$, then $\Lambda \gg 1$.}. If $x^A=(x')^A$, i.e. $dx^A=0$, the product \eqref{eprod3} reads
\be
\label{nomcd}
\langle e_\lambda(x) | e_\lambda(x) \rangle =  2 \Lambda.
\ee
This is the norm of the eigenvector $| e_\lambda(x) \rangle$; it diverges in the limit $\Lambda \rightarrow \infty$. This is the same divergence we have met discussing $e$-eigenvectors corresponding to real eigenvalues. As we recall,  regularized eigenvectors \eqref{ee2} have been constructed  multiplying by infinitesimal prefactor $1/ \sqrt{2 \Lambda}$. However, since now the eigenvectors are labeled by continuous variables, we will not do so. For instance, for the standard position operator eq. \eqref{dxpos} gives $\langle x | x \rangle = \delta^3(0)$. This is why, contrary to the analysis presented in the last section, we will not demand the position eigenvectors \eqref{eex} to be normalized. We only require they are orthogonal. 

The last translates into the additional constraint. Recalling the form  \eqref{eprod3} of the product, the constraint reads
\be
\label{ctheta}
 \Theta_{A B}(x) d x^A d x^B = -\Lambda^{-2}.
\ee
This is non-trivial only if $dx^A \neq 0$ (if $dx^A = 0$ the constraint is a non-issue because the product reduces to infinite norm \eqref{nomcd}). Assume for a moment that eq. \eqref{ctheta} has already been satisfied. One gets
\be
\label{tod}
 \langle e_{\lambda'}(x') | e_\lambda(x) \rangle =
    \begin{cases}
      \lim_{\Lambda \rightarrow \infty} 2 \Lambda, & \text{if}\ x=x' \\
      0, & \text{if}\ x \neq x'.
    \end{cases}
\ee
In the limit $\Lambda \rightarrow \infty$, one identifies $\delta (0) = \Lambda/(2 \pi)$. Therefore  eq. \eqref{tod} can be interpreted as related with Dirac delta, i.e. $\langle e_{\lambda'}(x') | e_\lambda(x) \rangle \propto  \delta(\lambda(x)-\lambda(x'))$. The operator $e(x)$ is  Hermitian in the resulting restricted Hilbert space and has continuous spectrum.  We should also keep in mind that eq. \eqref{tod} makes sense only in a close neighborhood of a fixed point. Hence, eq. \eqref{ctheta} is a local constraint.

\subsection{Spacetime interpretation of the constraint equation and geometry of $AdS_5$}
\label{constraint}

We will now interpret and solve the constraint \eqref{ctheta}. We start with an observation that linear combinations of $e$-operators give rise the standard creation and annihilation operators, leading to the standard algebra \eqref{alg2b} of the harmonic oscillator. Physically, this means that variables $x$ of $e$-operators can be identified with positions in space. Indeed, recalling standard interpretation of quantum fields, each point in space can be viewed as associated with a quantum harmonic oscillator, while the field itself can be interpreted as infinite number of such oscillators. The vacuum space can be related with oscillators in the ground state.

This simple logic can be further generalized to incorporate time. To be more specific, consider the possibility that the variables $x = x^0,...,x^{D-1}$ represent spacetime labels, just like for operators in the Heisenberg picture. However, in light of the algebra \eqref{alg2b}, this would mean there are oscillations also in temporal direction; something which requires the concept of operator of time. The latter is not a new idea and has already been discussed \cite{Susskind:1964zz,Bauer:2014qya,Gozdz:2019wbx}. One of the reasons behind such an approach is that time and space are treated on equal footing, as in the case of general relativity (or more generally, classical theory of gravity). Since on the quantum level time and space are represented by operators, this suggests a way of incorporating quantum gravity effects.

Despite operator of time is a promising idea, we will not go deeply into the subject. Instead, keeping in mind all what we have said, we restrict to say that the labels $x$ of $e(x)$ and $\partial(x)$ can be potentially connected with coordinates in spacetime. The way they are related should be, at least at the most superficial level, determined by solutions to the constraint \eqref{ctheta}.

Unfortunately, eq. \eqref{ctheta} cannot say much about the connection. This is because we actually did not discuss any quantum system. The latter would require introducing quantum fields, or performing a simplified analysis similar to the one presented in \cite{Plewa:2018llx}. Rather than searching for the connection between quantum states and geometry, we ask about the potential geometrical meaning of the constraint itself.

The starting point will be the result found in section \eqref{zeroeig}, a class of regularized eigenvectors spanning four-dimensional Minkowski space. As we recall, for $x=const$ and $a(x)=1$, eigenvectors \eqref{eex} are equivalent to the simple $e$-eigenvectors \eqref{eei}. Since for $\lambda=0$ the former was interpreted geometrically in terms of Minkowski space, the same is expected for the latter. Therefore, solving the constraint equation \eqref{ctheta}, one expects to find a solution sharing the same geometrical interpretation. This should hold for sufficiently small eigenvalues. In particular, we expect to get the four-dimensional Minkowski space in the limit $\lambda(x) \rightarrow 0$. We now examine this in details.

In order to do so, assume that $\lambda(x)$ vanishes at a fixed point $x^A=\vec{0}$, i.e. $\lambda(0)=0$. Obviously the point $x^A = \vec{0}$ is not special, this is only to illustrate the procedure. We also assume the function is slowly varying in a close vicinity of this point. This means that in the Taylor series approximating $\lambda(x)$ close to zero all gradient terms are small. In particular, $|\partial_A \lambda(0)|, |\partial_A \partial_B \lambda(0)| \ll 1$ and $|\partial_A \partial_B \lambda(0)| \ll |\partial_A \lambda(0)|$, etc. For a slowly varying function, eq. \eqref{ctheta} can be approximated as
\be
\label{ctheta0}
 \Theta_{A B}(0) d x^A d x^B = -\Lambda^{-2},
\ee
where 
\be
 \Theta_{A B}(0) = \frac{\partial_{A B} \lambda(0)}{6 \lambda(0)}.
\ee
Since the matrix $\Theta_{A B}(0)$ is symmetric, it can be diagonalized. Let $\Theta_{\mu \nu}^{(0)} = diag \{c_0,...,c_{D-1} \}$ stands for the result of diagonalization procedure. This procedure involves linear transformation of variables $x^A$. If the original matrix is already diagonal, they remain the same. If not, they will be changed. For simplicity we ignore such details now, assuming that $x^A$ refer to already diagonal form $\Theta_{\mu \nu}^{(0)}$. The constraint \eqref{ctheta0} reads
\be
\label{cetor}
\sum_{n=0}^{D-1} c_n (d x^n)^2 = -\Lambda^{-2}.
\ee
Here all (constant) coefficients $c_n$ are non-zero. In general, diagonalization of $ \Theta_{A B}(0)$ may result also in zero diagonal components $c_n$. We simply assume that $\lambda(x)$ was chosen is such a way that $\Theta_{\mu \nu}^{(0)}$ has $D$ non-zero components. Alternatively, one can reinterpret $D$ as representing the number of non-zero components after diagonalization.

Eq. \eqref{cetor} can be further simplified introducing the following new variables
\be
\tilde{x}^A  := \sqrt{|c_A|} \, x^A
\ee
(there is no summation convention here). The constraint \eqref{cetor} takes the form
\be
\label{cetor2}
\sum_{n=0}^{D-1} (\pm)_n (d \tilde{x}^n)^2 = -\Lambda^{-2},
\ee
where $(\pm)_n := \text{sign}(c_n)$.

We would like to examine if it is possible to get four-dimensional Minkowski space as a solution to the constraint. Obviously, this requires $D \geq 4$. In the simplest possible case $D=4$ while $\Theta^{(0)}_{A B}$ has a signature $(-,+,+,+)$. Recalling eq. \eqref{thetaform} one concludes that this will be the case for 
\be
\label{minform}
\lambda(x) = \lambda_0 \sin(\omega x^0)+ \lambda_0 \sinh(k_1 x^1)+\lambda_0 \sinh(k_2 x^2)+\lambda_0 \sinh(k_3 x^3),
\ee
where $\omega$, $k_1$, $k_2$, $k_3$ are assumed to be small to guaranty the function is slowly varying. Letting $l_0$ to be a dimensional length scale, the function will be slowly varying in the corresponding region if $\omega  \ll l_0^{-1}$ and $k_i  \ll l_0^{-1}$. The constraint \eqref{cetor2} reads
\be
\label{solut1}
\eta_{\mu \nu} d x^\mu d x^\nu = -\Lambda^{-2}.
\ee
As we recall, $\Lambda$ stands for a regularization parameter, which goes to infinity in the end. Geometrically, the left hand side of eq. \eqref{solut1} can be identified with the line element in four-dimensional Minkowski space, i.e. $ds^2 = \eta_{\mu \nu} d x^\mu d x^\nu$. In the limit $\Lambda \rightarrow \infty$ the latter goes to zero meaning that we cover not the whole space, but a hypersurface corresponding to the set of events $ds^2=0$. This can be identified with trajectories of massless particles. They define a hypersurface in $D=4$ Minkowski space, but not the whole space.

Fortunately, there is an even larger class of solutions. Notice that the right hand side of eq. \eqref{cetor2} is invariant under $SO(n,m)$ rotations, where $(n,m)$ stands for the signature of the matrix $\Theta^{(0)}_{A B}$. As such, eq. \eqref{cetor2} resembles the constraint defining the so-called pseudo-hyperbolic  space \cite{BADITOIU:2002nq} as embedding in $\mathbb{R}^{n,m}$. Depending on the signature of $\Theta^{(0)}_{A B}$ this includes hyperbolic, anti-de Sitter, and many other spaces with no clear spacetime interpretation\footnote{This is because more than one time-like direction and the presence of closed timelike curves which cannot be eliminated in a consistent way.}. Another observation is that eq. \eqref{cetor2} (as well as eq. \eqref{cetor}) becomes scale-invariant in the limit $\Lambda \rightarrow \infty$. The last two facts show that if anti-de Sitter space is a solution to the constraint, the solution is special.  Below we show that eq. \eqref{cetor2} indeed support that space and, in particular, $AdS_5$. 

In order to do so, assume that  $\lambda(x)$ is a function of six continuous variables such that the matrix $\Theta^{0}_{A B}$ has a signature $(-,-,+,+,+,+)$. The constraint \eqref{cetor2} reads
\be
\label{con1}
-(d \tilde{x}^0)^2 + (d \tilde{x}^1)^2 + (d \tilde{x}^2)^2 + (d \tilde{x}^3)^2 + (d \tilde{x}^4)^2- (d \tilde{x}^5)^2 = -\Lambda^{-2}.
\ee
Again, we can interpret the constraint \eqref{con1} as defining a hypersurface $ds^2=0$. However, we would not like to do so. Instead, we solve the constraint explicitly, installing a coordinate system on the resulting manifold. Note that the left hand side of eq. \eqref{con1} is invariant under $SO(4,2)$ rotations, the isometry group of $AdS_5$. Hence, we expect that anti-de Sitter space is a solution to the constraint \eqref{con1}. More precisely, let  $\delta L:= \Lambda^{-1}$ and
\be
\delta x^A := dx^A
\ee 
stand for components of an infinitesimal, normalized vector in $\mathbb{R}^{4,2}$. The normalization is provided by the constraint \eqref{con1}:
\be
\label{con2}
- (\delta x^0)^2+ (\delta x^1)^2 +...+ (\delta x^4)^2 - (\delta x^5)^2 = -\delta L^2.
\ee
The left and the right hand side of the equation above are built out of infinitesimally small objects of the same order. A convenient way of solving the constraint \eqref{con2} is installing Poincare coordinate system $(t_\delta,{x_\delta}_1,{x_\delta}_2,{x_\delta}_3,{z_\delta})$ \cite{Bayona:2005nq}, defined as
\begin{align}
\nonumber
\delta x^0 &= \frac{-t_\delta^2+{x_\delta}_1^2+{x_\delta}_2^2+{x_\delta}_3^2+z_\delta^2+\delta L^2}{2 z_\delta},
\\[1ex]
\nonumber
\delta x^i &= \delta L \frac{{x_\delta}_i}{z_\delta}, \quad i=1,2,3,
\\[1ex]
\nonumber
\delta x^4 &= \frac{-t_\delta^2+{x_\delta}_1^2+{x_\delta}_2^2+{x_\delta}_3^2+z_\delta^2-\delta L^2}{2 z_\delta},
\\[1ex]
\delta x^5 &= \delta L \frac{t_\delta}{z_\delta}.
\label{xcdr}
\end{align}
Inserting eqs. \eqref{xcdr} into eq. \eqref{con2} one checks the constraint is satisfied. Note that unlike the standard construction of $AdS_5$ as embedding in $\mathbb{R}^{4,2}$, eq. \eqref{con2} is a constraint on a vector $\delta x^A \in \mathbb{R}^{4,2}$ built out of infinitesimal variables. Solutions \eqref{xcdr} lead to the following metric on the resulting manifold
\be
\label{met}
ds^2 = \delta L^2 \frac{-dt_\delta^2+d{x_\delta}_1^2+{dx_\delta}_2^2+{dx_\delta}_3^2+dz_\delta^2}{z_\delta^2}.
\ee
This is the metric of $AdS_5$ in Poincare coordinates; $\delta L$ stands for AdS radius.
Formally, the right hand side of eq. \eqref{met} is small of order four. This is actually not a problem since the line element is manifestly scale-invariant. Small lengths can be isomorphically map into their arbitrarily large equivalents. In particular, the line element \eqref{met} is invariant under scaling transformation $(t_\delta, x_{\delta_1}, x_{\delta_2}, x_{\delta_3}, z_{\delta}) \rightarrow (\Omega t_\delta, \Omega x_{\delta_1}, \Omega x_{\delta_2}, \Omega x_{\delta_3}, \Omega z_{\delta})$, where $\Omega$ is arbitrary real parameter. This scale-invariance  reflects the scale-invariance of the constraint \eqref{con2} or, alternatively, the form \eqref{ctheta0}. 

Since $\delta L = \Lambda^{-1}$, AdS-radius goes to zero in the limit $\Lambda \rightarrow \infty$, while  the scalar of curvature  $R = -20 \Lambda^2 = -20 \, \delta L^{-2}$ becomes infinite. Fortunately, this is not an issue at the conformal bound, corresponding to $z_\delta = 0$ in Poincare coordinates. To be more precise, taking the limit $z_\delta \rightarrow 0$ leads to the ''boundary'' of $AdS_5$. This turns out to be nothing but the four-dimensional Minkowski space:
\be
\label{bound}
ds^2 = -dt^2+d{x}_1^2+{dx}_2^2+{dx}_3^2.
\ee
The metric above can be easily derived from eq. \eqref{met}, assuming that $\delta L, z_\delta \rightarrow 0$ while keeping $\delta L / z_\delta =const$. Letting $z_\delta \rightarrow 0$ is a way to compensate the problematic limit $\delta L \rightarrow 0$. The five-dimensional geometry effectively reduces to the four-dimensional one, and the letter is manifestly non-singular. It takes the form of the four-dimensional Minkowski space. However, the space emerges in a rather unexpected way as the boundary of a higher-dimensional spacetime. 

At this point it is worth underlying that all what we have said concerns the assumption that $\lambda(x)$ is small and slowly varying. What is more, we restricted ourselves to the local version of the constraint equation, valid in a close neighborhood of a fixed point. Even with these strong assumptions we arrived with a large class of solutions, identifying $AdS_5$ to be special because of the form of its boundary. However, global solutions to the constraint equation can be qualitatively different. Without specifying the form of the function $\lambda(x)$ we cannot say much about them. The only what we know is that locally they should look like $AdS_5$ to match the results found for $e$ operator.  Quite interestingly, the last fact can be viewed as a holographic analogue of the standard local flatness. The Minkowski space emerges in a holographic manner, being the boundary of the anti-de Sitter space.

We close the analysis with a simple  observation regarding the form of the product \eqref{eprod2}. It is straightforward to observe that the latter is invariant under scaling transformations $\lambda(x) \rightarrow b \lambda(x)$. This means that scale-invariance is a symmetry of the product in its most general form. In the spacial case, restricting to small regions in the parameter space and small $\lambda(x)$, this symmetry is reflected by scale-invariance of $AdS_5$.

\subsection{Momentum eigenequation}

So far we considered $e$-position, represented either by $e$ or $e(x)$ operators. Now we will discuss the momentum. In the simplest possible case, it reads $\hat{p}=-i \partial$. The corresponding eigenequation
\be
\hat{p} | p \rangle = \lambda |p \rangle,
\ee
has a solution
\be
\label{mom}
|p\rangle =  \hat{P}_{\lambda} |\phi \rangle,
\ee
where
\be
\hat{P}_{\lambda} = \exp(i \lambda e) +\sum_{i=1}^{\infty} \frac{\partial^n}{(i \lambda)^n}.
\ee
The scalar product of two eigenvectors \eqref{mom} reads
\be
\label{mom2}
\langle p_{\lambda_i} | p_{\lambda_j} \rangle = 1+\sum_{n=1}^\infty \left[ \left(\frac{\lambda_i^*}{\lambda_j}\right)+\left(\frac{\lambda_j}{\lambda_i^*}\right) \right].
\ee
Notice that the right hand side of eq. \eqref{mom2} is the same as the right hand side of eq. \eqref{Eij}, found for $e$-position. Therefore, the future analysis, including construction of normalized eigenvectors and their geometric interpretation, will be also the same. The superselection rules guaranty the momentum is conserved, while the whole $D=4$ Minkowski space corresponds to the zero momentum eigenvalue. Again, both spacetime dimension and signature of the metric are determined by the regularization procedure.

For generalized momentum  $-i \partial(x)$ the eigenequation takes the form
\be
\label{peig}
-i \partial(x) | p(x) \rangle = \lambda(x)| p(x) \rangle.
\ee
The eigenvector can be found writing down the most general form of the state in the maximally extended space:
\begin{align}
\nonumber
| p(x) \rangle  = \Big( a(x) + \sum_{n=1}^\infty \int d^D x_1 ... d^D x_n \, b_n(x;x_1,...,x_n) e(x_1) ...  e(x_n) +
\\[1ex]
+\sum_{n=1}^\infty \int d^D x_1 ... d^D x_n \, c_n(x;x_1,...,x_n) \partial(x_1)  ...  \partial(x_n) \Big)|\phi \rangle.
\label{pe}
\end{align}
Substituting eq. \eqref{pe} into \eqref{peig} and solving the resulting equations gives
\begin{align}
&b_n(x;x_1,...,x_n) = a(x) \frac{i^n}{n!} \lambda^n(x),
\\[1ex]
&c_n(x;x_1,...,x_n) = a(x) \frac{(-i)^n}{\lambda^n(x)} \prod_{k=1}^n \delta(x-x_k).
\end{align}
Here $a(x)$ is a remaining functional parameter. The product of two eigenvectors \eqref{pe} reads
\be
\label{peprod}
\langle p(x')| p(x) \rangle = a^*(x') a(x) \left( 1+ \sum_{n=1}^\infty \left[  \left( \frac{\lambda(x)}{\lambda^*(x')}\right)^n + \left( \frac{\lambda^*(x')}{\lambda(x)}\right)^n \right]  \right).
\ee
Again, we found the same result as for the extended position (see eq. \eqref{eprod}). Therefore, the constraint equation will be also the same. Having found $D=4$ Minkowski space discussing $e$-position, the same can be found discussing $e$-momentum. The space emerges as a conformal boundary of $AdS_5$.

\subsection{The most general operators}

We are ready to generalize the results to the case of the most general $e$-operators, $e_a(x)$ and $\partial_a(x)$, $a=1,...,d$. The position eigenequation reads
\be
\label{egei}
e_a(x) | e (x) \rangle = \lambda_a(x) | e (x) \rangle,
\ee
where
\be
\label{xgen}
 | e (x) \rangle = \bigotimes_{a=1}^d |e_a (x) \rangle,
\ee
and $|e_a (x) \rangle$ stand for eigenvectors \eqref{eex}. There are $d$ copies of such eigenvectors, one for each label $a$. It is straightforward to observe that the rules \eqref{init}-\eqref{gchain2} guaranty that eigenvectors $|e_a(x)\rangle$ indexed by different discrete labels are orthogonal. 

Notice that the solution \eqref{xgen} is similar to the previously found \eqref{eex}. In fact, the only what changed is that now we have $d$ functional eigenvalues $\lambda_a(x)$, while eigenvectors  \eqref{xgen} were built out of ''simple'' eigenvectors \eqref{eex}. We have a copy of the constraint equation \eqref{ctheta} for each discrete label of the $e$-position.  

It is convenient adopting a bit more natural notation, rewriting the eigenequation \eqref{egei} as
\be
\label{egei2}
\hat{X}^a(x) | X(x) \rangle = X^a(x) | X(x) \rangle.
\ee
Here we defined
\be
\hat{X}^a(x) := e_a(x), \quad X^a(x):= \lambda_a(x), \quad | e (x) \rangle:=| X(x) \rangle.
\ee
Similarly, for the momentum $\hat{P}^a(x) := -i \partial_a(x)$, one has
\be
\label{pgei}
\hat{P}^a(x) | P(x) \rangle = P^a(x) | P(x) \rangle.
\ee
The operators $\hat{X}^a(x)$, $\hat{P}^a(x)$ are similar to the standard position and momentum, however, do not share the standard algebra. Instead, eq. \eqref{genxpalgebra2} can be rewritten as
\be
\label{XP}
[\hat{X}^a(x),\hat{P}^b(y)] = i \delta_{a b} \delta^D(x-y)\hat{\eta}.
\ee
As it was shown in \cite{Plewa:2018llx}, eigenstates of creation and annihilation operators can be constructed in such a way they belong to extended, but not maximally extended space. This means that one can skip the operator $\hat{\eta}$, reproducing the standard algebra of the quantum harmonic oscillator. However, for $e$-position and $e$-momentum the situation is qualitatively different. Solving eigenequations one cannot skip the term responsible for the presence of the bare fiducial vector. In consequence, the commutation relation cannot be simplified in the space of eigenvectors. Therefore we cannot expect the standard uncertainty relation for these operators.

Discussing eigenequation of $e(x)$ operator we identified the constraint \eqref{ctheta} guarantying that  eigenvectors from different eigenvalues are orthogonal. Itself, the constraint was interpreted geometrically as defining spacetime as embedded manifold. Letting $D=6$ we found $AdS_5$ as embedding in $\mathbb{R}^{4,2}$. What makes the solution special is that it matches perfectly the result found for ''simple'' $e$-operators: the eigenspace corresponding to the zero eigenvalue can be interpreted geometrically as four-dimensional Minkowski space. As we recall, this was generalized to the case of generalized $e$-operators in the limit of small $\lambda(x)$. Keeping this in mind, it is natural to ask what will happen in the most general case, i.e. for $e_a(x)$ and $-i \partial_a(x)$. 

As before, we start with $e$-position. Due to the symmetry of the scalar product, the same results can be found for the momentum. In case of $e_a(x)$ eigenstates, one finds $d$ copies of the constraint equation \eqref{ctheta}:
\be
\label{cnstr1}
\Theta^a_{A B}(x) d x^{A} d x^{B} = -\Lambda^{-2}_a,
\ee
where
\be
\label{Th}
\Theta^a_{A B}(x) = \frac{\partial_{A B}X^a(x)}{6 X^a(x)},
\ee
and $a=1,...,d$. Following the construction presented in section \ref{constraint} we will be interested in covering a small region where the eigenvalues are close to zero. The latter is defined by a neighborhood of $X^a=0$. Suppose that $X^a(0)=0$. Repeating the steps of the construction outlined in section \ref{constraint}, we assume that each $X^a(x)$ is a slowly varying function, rewriting the constraints \eqref{Th} as  
\be
\label{cetorA}
\sum_{n=0}^{D-1} c_n^a (d x^n)^2 = -\Lambda^{-2}_a.
\ee
Here $\Lambda_a$ stand for regularization parameters. Without loosing of generality one can assume they are all equal, i.e. $\Lambda_1=...=\Lambda_d = \Lambda$. Again, one should take the limit $\Lambda \rightarrow \infty$ in the end. Introducing new variables
\be
\tilde{x}^{a,A} := \sqrt{|c_A^a|} \, x^A
\ee
(there is no summation convention here), one rewrites eq. \eqref{cetorA} as
\be
\label{cetor2A}
\sum_{n=0}^{D-1} (\pm)_n^{(a)} (d \tilde{x}^{a,n})^2 = -\Lambda^{-2}.
\ee
Here $(\pm)_n^{(a)} := \text{sign}(c_n^a)$. Suppose we have chosen $X^a(x)$ in such a way that each matrix $\Theta^a_{A B}$ has a signature $(-,-,+,+,+,+)$. This would be the case for
\be
\label{xa}
X^a(x) = \lambda^a_0 \sin(\omega^a x^{0}) + \lambda^a_0  \sum_{i=1}^{3} \sinh(k_i^a x^{i}) + \lambda^a_0  \sin(k_4^a x^{4}),
\ee
where $\lambda^a_0$, $\omega$, $k_1^a,...,k_4^a$ are additional parameters. As before, we assume that $\omega$, $k_1^a,...,k_4^a$ are small with respect to some length scale $l_0$: $\omega^a  \ll l_0^{-1}$, $k_i^a  \ll l_0^{-1}$, $k_4^a  \ll l_0^{-1}$. This guaranties $X^a(x) $ are slowly varying. 

Following the construction of $AdS_5$ in Poincare coordinates presented in section \ref{constraint} one concludes that starting with the form \eqref{xa}, one finds $d$ copies of that space, one for each discrete label $a$. Since each time one gets the same geometric result, one can start with a simpler form of $X^a(x)$:
\be
\label{xa0}
X^a(x) = \lambda_0^a \sin(\omega x^{0}) + \lambda_0^a  \sum_{i=1}^{3} \sinh(k_i x^{i}) + \lambda_0^a  \sin(k_4 x^{4}),
\ee
where $\omega$, $k_1,...,k_4$ are the same for all discrete labels $a$. It is straightforward to check that one finds the same constraint for each discrete label $a$. Geometrically, this results in $AdS_5$. This is justified within the adopted assumption that eigenvalues are small, i.e.
\be
\label{Xsmall}
| X^a(x)| \ll 1.
\ee
We should keep in mind that all we have said concerns the local form of the constraint given by eq. \eqref{ctheta0}. Therefore, the resulting geometry is only a local solution. This is somehow similar to local flatness in general relativity. Note that we have already mentioned this analogy discussing eigenvectors of $e(x)$ operator. As previously, general solutions to the constraint \eqref{ctheta} may differ significantly from anti-de Sitter or even Minkowski space. Unfortunately, we canot say much about them without incorporating matter (quantum fields, vibrations of strings, etc.).

\section{Conclusions}
\label{discussion}

We have discussed the solutions to the eigenequation of the $e$-operators, the position and momentum.
Starting with the eigenvectors of the $e$ operator, we show  that they are non-normalizable, however,
can be easily regularized. We observe that regularized eigenvectors span restricted (extended) Hilbert
space in the form of a direct sum of separate spaces of finite dimension. This includes two-dimensional
spaces spanned by eigenvectors of  opposite eigenvalues, and a four-dimensional space composed of
degenerated states corresponding to $\lambda=0$. We also identified superselection rules forbidding
combinations of states with different absolute values of the eigenvalues $\lambda$. Our construction
guaranties that operator the  $e$ is Hermitian if it is restricted to any of these spaces. The same
result holds for $e$-momentum. The space corresponding to the zero eigenvalue turns out to be special,
because it has the maximal dimension and because the orthonormal basis consists of three positive and
one negative eigenvector. As such, the space can be identified with the four-dimensional Minkowski space.
The interpretation relies upon the observation that for $\lambda=0$ any superposition of eigenvectors
is also an eigenvector corresponding to the same zero eigenvalue, and amplitudes in the superposition
are linked with spacetime coordinates. The difference is that instead of Lorentz symmetry group, we
arrive with a more general group of complex Lorentz rotations $O(3,1,\mathbb{C})$. Rewritten in terms
of real parameters $\theta^\mu$ and $\sigma^\nu$, the rotations result in non-trivial transformations
given by eqs. \eqref{tr1}--\eqref{tr2}. Assuming that half of the directions in the parameter space
are ``frozen'', i.e. either $\sigma^\mu = \sigma^\mu_0 =const$ or $\theta^\mu = \theta^\mu_0 =const$,
leads to the standard four-dimensional Poincare group.

Having found eigenvectors of the simplest $e$-operators, we generalize the result to the case of
operators with continuous labels. Considering the position eigenequation, we notice the presence of
the additional constraint equation \eqref{ctheta} which ensures that the operator is Hermitian.
Eigenvalues $\lambda(x)$ turned out to be labeled by the labels of the $e$-operators, and except
eq. \eqref{ctheta} there was no further constraints on the form of the function $\lambda(x)$.
We discuss  the constraint equation \eqref{ctheta} restricting to the region where the eigenvalue
is close to zero. This is supplemented by technical assumption making $\lambda(x)$ slowly varying
function. The reason for considering small eigenvalues was dictated by the observation that for
the simplest $e$-operators, the space corresponding to the zero eigenvalue was identified with
the four-dimensional Minkowski space. In what follows, taking $\lambda(x)$ to be close to zero,
we expect to reproduce the geometric result corresponding to $\lambda=0$, i.e. $D=4$ Minkowski
space. We show that this is indeed the case and in a class of various possible solutions to the
constraint, one can identify the one interpreted geometrically as $AdS_5$. The Minkowski space
emerges in a holographic way, as the conformal boundary of the resulting anti-de Sitter space.

Finally, we discuss the eigenequation for the most general $e$-operators, labeled both by discrete
and continuous labels. In particular, for the position operator $e_a(x)$, we found $d$
constraint equations, where $d$ stands for the number of discrete labels of the operator.
We show that solutions to the constraints in the region of small eigenvalues also supports
$AdS_5$. We notice  an interesting symmetry between $e$-position and $e$-momentum, consisting
in the structure of the scalar product. Taking a closer look at the position eigenequation
\eqref{egei} and rewriting it in the form \eqref{egei2}, we find an analogy to string theory:
eigenvalues $X^a(x)$ resemble worl-volume coordinates. The latter can be seen as a starting
point for string theory. Indeed, on a classical level this provides motivation for writing
down the standard p-brane action. However, specifying the classical field $X^a(x)$,
one gets automatically the corresponding quantum state $|X^a(x)\rangle$ in the maximally
extended space.

The main result presented in this paper is the geometric interpretation of solutions to
the eigenequation of $e$-operators. The interpretation relies upon the number and sign
of the norm of regularized eigenvectors corresponding to the zero eigenvalue. In the
simplest case this gives rise the four-dimensional Minkowski space. For more general
operators the space emerges as the boundary of $AdS_5$. It is also worth underlying
that for simple operators, $e$ and $\partial$, there is no direct link between classical
coordinates and their quantum equivalents. The whole Minkowski space corresponds to
a single position or momentum eigenvalue, while superselection rules guaranty these
quantities are conserved. In the case of general $e$-operators the connection becomes
slightly more transparent. Still, similarly as in \cite{Plewa:2018llx}, spacetime
turns out to be a secondary concept, associated with quantum states.

Compared to the discussion presented in \cite{Plewa:2018llx,VanRaamsdonk:2009ar,VanRaamsdonk:2010pw},
the quantum entanglement seems not to be essential for the emergence of spacetime. In what follows,
one can ask how it is possible we extracted classical geometry without incorporating quantum
correlations. The answer is that these correlations are actually present, being hidden in the
adopted assumptions. More specifically, the interpretation of the continuous labels of $e$-operators
bases on the algebra \eqref{alga} and its connection with the quantum harmonic oscillator. We have
also utilized the fact that quantum fields can be interpreted as infinite number of oscillators
associated with points in space. This holds true both for excited states and the ground state.
The latter is particularly interesting since the vacuum state of relativistic QFT is maximally
entangled \cite{Harlow:2014yka}. Assuming that empty space is filled with quantum fields in the
lowest energy state, we get a link with entanglement. In particular, the vacuum anti-de Sitter
can be seen as represented by infinite number of oscillators in the ground state.

Talking about the geometry of anti-de Sitter space, it is worth mentioning that in string
theory the latter appears when discussing gravitational field of D-branes. In the {\it closed
string description}, they correspond to spacetime geometry in which close strings propagate.
In particular, in the case of type IIB supergravity the spacetime metric sourced by a system
of $N$ branes, in the strong gravity region, reduces to $AdS_5 \times S_5$
\cite{CasalderreySolana:2011us,Gibbons:1987ps,Garfinkle:1990qj,Horowitz:1991cd}.
On the other hand, D-branes themselves are objects where endpoints of open strings are
attached. Consequently, in the {\it open string description}, D-branes can be viewed
as sort of hypersurfaces. Their excitations are open strings living on the brane,
and closed strings propagating in the bulk \cite{CasalderreySolana:2011us,Hubeny:2014bla}.
The dual description of D-branes is essential for AdS/CFT conjecture, leading to a
link between supersymmetric conformal Yang-Mills theory at the boundary and gravity
in asymptotically anti-de Sitter space in the bulk.

Notice that we have met a brane-like analogy two times in our paper. Firstly, discussing
the symmetry group of complex rotations $O(3,1,\mathbb{C})$, we show they lead to the
standard Poincare symmetry in four dimensions if the initial symmetry is partially broken.
This is interpreted as defining  a four-dimensional hypersurface in higher dimensional space,
something which resembles a brane. Second, for the most general operators the position
eigenvalues take the form of a worldvolume in the sense that there is a map $X^a(x)$,
i.e. $x^A \mapsto X^a$. These functions are not fixed by the construction, however,
solving the constraint \eqref{ctheta0} we assume they are small and slowly varying.
For a brane this would mean that we are in the region where perturbations are small.
Quite interestingly, recalling standard results of the AdS/CFT correspondence, this
corresponds to the geometry dual to strongly-coupled Yang-Mills plasma close to thermal
equilibrium \cite{Booth:2010kr,Plewa:2013rga}. The main difference is that in our case
the AdS-radius goes to zero, making spacetime curvature in the bulk divergent. Therefore,
trying to interpret our resulting geometry as being sourced by a brane, we should keep
in mind we are in the region where AdS radius is small. A connection to D-branes in string theory consist in the fact that eigenvalues of the most general $e$-operators support worldvolume interpretation. Moreover, as in case of D-brane system, we identified the geometry of anti-de Sitter space. The latter is a special solution to the constraint, reflecting its scale-invariance.

It would be interesting to ask if such an interpretation is more than
a formal analogy to string theory. In fact, the formalism of the $e$-operators in
the extended Hilbert space was already used in \cite{Plewa:2018llx} while searching
for the connection between the ground state of a two-dimensional oscillator and $AdS_3$.
Again, this is in full consistency with the holographic principle. All of these facts
suggest that Hilbert space with no positive-definite scalar product can be a useful
concept combining  quantum mechanics and gravity.

The construction presented in this paper can be further generalized in two different ways.
The first one is to examine an explicit identification of the position eigenvalues $X^a(x)$
with coordinates on the worldvolume of a brane. That is, one can consider the possibility
of letting arbitrary functions $X^a(x)$ to be solutions of the classical equation of motion
for a p-brane. Alternatively, one could formally introduce quantum fields, expressing them
in terms of the $e$-operators. The background geometry will be provided by the solution to
the constraint, however, one expects a further backreaction induced by quantum correlations.
This, in fact, could be the first step towards identifying a more general class of
gravitational dualities.

\bigskip

\acknowledgments I would like to thank Piotr H Chankowski and W\l{}odzimierz Piechocki for helpful discussions. 

\appendix

\section{Regularization}
\label{cal}

We consider the product $\langle e_{\lambda_i}|e_{\lambda_j} \rangle = \langle \phi | (E_{\lambda_i})^\dagger E_{\lambda_j} | \phi \rangle$ of two eigenvectors \eqref{eei}, labeled by two eigenvalues $\lambda_i$ and $\lambda_j$. Denote $\langle \phi|...|\phi \rangle := \langle ... \rangle$, skipping the fiducial vector for convenience. The product reads
\begin{align}
\nonumber
\langle e_{\lambda_i}|e_{\lambda_j} \rangle =  \langle  E_{\lambda_i}^\dagger E_{\lambda_j}  \rangle =  \bigg \langle \bigg(  1+\sum_{k=1}^{\infty} \frac{e^k}{(\lambda_i^*)^k} + \frac{(\lambda_i^*)^k}{k!} \partial^k\bigg)
  \bigg(  1+\sum_{n=1}^{\infty} \frac{e^n}{\lambda_j^n} + \frac{(-\lambda_j)^n}{n!} \partial^n \bigg)  \bigg \rangle,
\label{expo1}
\end{align}
where in the last equality we substituted the exact form of the operator $E_\lambda$ \eqref{Eop}, making use eqs. \eqref{eherm} and \eqref{dherm}. Multiplying terms on the right hand side of eq. \eqref{expo1} gives
\begin{align}
\nonumber
 \langle  E_{\lambda_i}^\dagger E_{\lambda_j}  \rangle &= 1 + \sum_{k=1}^{\infty} \sum_{n=1}^{\infty} \bigg( \frac{(-\lambda_j)^n}{(\lambda_i^*)^k} \frac{\langle e^k \partial^n \rangle}{n!} + \frac{(\lambda_i^*)^k}{\lambda_j^n} \frac{\langle \partial^k e^n \rangle}{k!} \bigg) =
\\[1ex]
&=1+ \sum_{n=1}^\infty \bigg( \left( \frac{\lambda_j}{\lambda_i^*}\right)^n +  \left( \frac{\lambda_i^*}{\lambda_j}\right)^n \bigg).
\end{align}
To get the first line, we used eq. \eqref{zera}, i.e. $\langle e^n \rangle = \langle \partial^n \rangle = 0$ (for $n>0$). To get the second, we utilized eqs. \eqref{diffeo1}-\eqref{diffeo2}, finding 
\be
\label{kron}
\langle  e^k \partial^n \rangle = (-1)^n \delta_{n k} n!, \quad \langle \partial^n e^k \rangle = \delta_{n k} k!
\ee
We are now ready to consider the product \eqref{crule}:
\begin{align}
\nonumber
: \langle  (E_{\lambda_i}^{\Lambda_i})^\dagger E_{\lambda_j}^{\Lambda_j} \rangle : \, \, \, =  
\frac{1}{2} \langle (E_{\lambda_i}^{\Lambda_i})^\dagger E_{\lambda_j}^{\Lambda_j}  \rangle \Big |_{\Lambda_i \leq \Lambda_j}  + \frac{1}{2} \langle  (E_{\lambda_i}^{\Lambda_i})^\dagger E_{\lambda_j}^{\Lambda_j}  \rangle \Big |_{\Lambda_i \geq \Lambda_j}.
\label{scrule}
\end{align}
For the first term, one finds
\begin{align}
\langle  (E_{\lambda_i}^{\Lambda_i})^\dagger E_{\lambda_j}^{\Lambda_j}  \rangle \Big |_{\Lambda_i \leq \Lambda_j} &= 
1 + \sum_{k=1}^{\Lambda_i} \sum_{n=1}^{\Lambda_j} \bigg( \frac{(-\lambda_j)^n}{(\lambda_i^*)^k} \frac{\langle e^k \partial^n \rangle}{n!} + \frac{(\lambda_i^*)^k}{\lambda_j^n} \frac{\langle \partial^k e^n \rangle}{k!} \bigg) =
\\[1ex]
&=1+ \sum_{n=1}^{\Lambda_i} \bigg( \left( \frac{\lambda_j}{\lambda_i^*}\right)^n + \left( \frac{\lambda_i^*}{\lambda_j}\right)^n \bigg).
\label{part1}
\end{align}
In the last line we utilized the fact that, according to eqs. \eqref{kron}, the only non-zero contribution to the sum comes from expectation values involving equal number of $e$-operators $e$ and $\partial$. In what follows, $k=n$. This also means that for $\Lambda_i \leq  \Lambda_j$ the summation $\sum_{n=1}^{\Lambda_j}(...)$ can be terminated at $\Lambda_j=\Lambda_i$. A similar analysis shows that
\be
\langle  (E_{\lambda_i}^{\Lambda_i})^\dagger E_{\lambda_j}^{\Lambda_j}  \rangle \Big |_{\Lambda_i \geq \Lambda_j} = 1+ \sum_{n=1}^{\Lambda_j} \bigg( \left( \frac{\lambda_j}{\lambda_i^*}\right)^n +  \left( \frac{\lambda_i^*}{\lambda_j}\right)^n \bigg).
\label{part2}
\ee
Substituting expectation values \eqref{part1}-\eqref{part2} into \eqref{scrule} gives
\begin{align}
\nonumber
: \langle  (E_{\lambda_i}^{\Lambda_i})^\dagger E_{\lambda_j}^{\Lambda_j} \rangle : \,\,\,  &=  1 + \frac{1}{2} \sum_{n=1}^{\Lambda_i} \bigg( \left( \frac{\lambda_j}{\lambda_i^*}\right)^n + \left( \frac{\lambda_i^*}{\lambda_j}\right)^n \bigg) +  \frac{1}{2} \sum_{n=1}^{\Lambda_j} \bigg( \left( \frac{\lambda_j}{\lambda_i^*}\right)^n +  \left( \frac{\lambda_i^*}{\lambda_j}\right)^n \bigg) = 
\\[1ex]
& = \frac{1}{2} \langle  (E_{\lambda_i}^{\Lambda_i})^\dagger E_{\lambda_j}^{\Lambda_i} \rangle + \frac{1}{2} \langle  (E_{\lambda_i}^{\Lambda_j})^\dagger E_{\lambda_j}^{\Lambda_j} \rangle.
\label{helul}
\end{align}
Taking into account eq. \eqref{helul}, the product \eqref{cel} reads
\begin{align}
\nonumber
: \langle e_{\lambda_i}^{(\Lambda_i)}|e_{\lambda_j}^{(\Lambda_j)} \rangle :  \,\,\, &=  1 + \frac{1}{2} \lim_{N_i \rightarrow \infty} \sum_{n=1}^{\Lambda_i(N_i)} \bigg( \left( \frac{\lambda_j}{\lambda_i^*}\right)^n + \left( \frac{\lambda_i^*}{\lambda_j}\right)^n \bigg) + 
\\[1ex]
&+ \frac{1}{2} \lim_{N_j \rightarrow \infty} \sum_{n=1}^{\Lambda_j (N_j)} \bigg( \left( \frac{\lambda_j}{\lambda_i^*}\right)^n + \left( \frac{\lambda_i^*}{\lambda_j}\right)^n \bigg).
\label{orel}
\end{align}
In particular, for the vectors \eqref{imvec}, one has\footnote{Do not confuse the imaginary number $i$ with indexes of $N_i$, $\Lambda_i$.} $\lambda_i = \lambda_j = i$, $\Lambda_i = 2 N_i$, $\Lambda_j = 2 N_j-1$, and
\begin{align}
\nonumber
\langle e_i^{(\Lambda_1)}|e_i^{(\Lambda_2)} \rangle &= 1 + \frac{1}{2} \lim_{N \rightarrow \infty} \sum_{n=1}^{2 N} 2 (-1)^n + \frac{1}{2} \lim_{N \rightarrow \infty} \sum_{n=1}^{2 N-1} 2 (-1)^n =
\\[1ex]
&= 1+ \lim_{N \rightarrow \infty} \Big[ \frac{1}{2} \left(  -1 + (-1)^{2 N} \right) + \frac{1}{2} \left(  -1 + (-1)^{2 N-1} \right)\Big]=0.
\end{align}

\section{Complex eigenvalues}
\label{cal2}
Below we discuss normalization of eigenvectors of $e$ operator corresponding to general complex eigenvalues $\lambda = r e^{i \varphi}$, where $r>0$ and $\varphi \in (0,\pi) \cup (\pi,2 \pi)$. Here we excluded two possibilities, $\varphi = 0$ and $\varphi = \pi$, since both they correspond to real eigenvalues  discussed in section \ref{re} (they require a different procedure). Define
\be
\label{imgen}
|e_\lambda^{(\Lambda)} \rangle  :=  \lim_{N \rightarrow \infty} E_\lambda^{\Lambda (N)} | \phi \rangle.
\ee
Let $L_\lambda[\Lambda] := \langle  (E_{\lambda}^{\Lambda})^\dagger E_{\lambda}^{\Lambda} \rangle$. In the limit $N \rightarrow \infty$ this is norm of the vector \eqref{imgen}. For $\lambda = r e^{i \varphi}$, one finds
\begin{align}
L_\lambda[\Lambda] = 1 + 2 \sum_{n=1}^{\Lambda} \cos(2 n \varphi) = 1 + 2 \sin^{-1}(\varphi) \sin(\Lambda \varphi) \cos \left((\Lambda+1)\varphi \right).
\label{creg1}
\end{align}
Clearly, the latter is ill-defined in the limit $\Lambda \rightarrow \infty$. The idea is to replace the cut-off $\Lambda$ by a map $\Lambda(N)$ such that the product \eqref{creg1} will be well defined in the limit $N \rightarrow \infty$. More precisely, we promote the cut-off $\Lambda$ to a one-to-one function
\be
\label{map}
\Lambda \mapsto \Lambda(N).
\ee
We expect that its form should be specified by the regularization procedure. We now discuss general conditions guarantying this to be the case. First,  we require that at least from some fixed point $N_0 \gg 1$, $ \Lambda(N)$ is increasing function of $N$. Otherwise, the replacement \eqref{map} would make no sense. Second, since we are interested in the limit $N \rightarrow \infty$, looking for a candidate for $\Lambda(N)$ it is sufficient to consider an asymptotic form valid for $N \gg 1$. One can take the following ansatz
\be
\label{asym}
\Lambda(N) = \sum_{k=0}^{N_{max}} a_k N^k, 
\ee
where $N_{max}$ stands for a cut-off, while $a_k$ are real coefficients chosen so that the function \eqref{asym} is increasing. Notice that the original form \eqref{creg1} is a $C^{\infty}$ function and, in particular,
\be
\label{prop1}
| \, \lim_{\Lambda \rightarrow \infty} \frac{d^k}{d \Lambda^k} L_\lambda [\Lambda] \, | < \infty 
\ee
for $k \in \mathbb{N}$. In what follows, if the regularized function $ L_\lambda [\Lambda(N)]$ has a well-defined limit, we require it goes smoothly in the limit, i.e. 
\be
\label{prop2}
| \, \lim_{N \rightarrow \infty} \frac{d^k}{d N^k} L_\lambda [\Lambda(N)] \, | < \infty. 
\ee
Recalling the form \eqref{creg1} of the norm, one concludes that the only way of how eq. \eqref{prop2} can be satisfied is restricting to a class of linear functions $\Lambda(N) = q N +l$, where $q>0$. Otherwise, the derivatives will become divergent as $N \rightarrow \infty$. Coefficients $a$, $b$ can be fixed requiring the function $L_\lambda(N)$ has a well-defined limit $N \rightarrow \infty$. It turns out, this will be the case for
\begin{align}
\label{regul1}
&\varphi = \frac{p}{q} \pi, \quad p,q \in \mathbb{N},
\\[1ex]
&\Lambda_{q,l}(N) = q N +l.
\label{regul2}
\end{align}
We now look closer the conditions \eqref{regul1}-\eqref{regul2}. Before doing so, however, we have to make some comments. The first observation is that in eq. \eqref{regul1} we should additionally assume that  $p \neq 0$ and  $p/q \neq 1/2$. This is to eliminate the possibility $\lambda \in \mathbb{R}$, making the right hand side of eq. \eqref{creg1} divergent. Without loosing of generality, one can also assume that
\be
p < 2 q,
\ee
as a consequence of the fact that $\varphi < 2 \pi$ (for complex numbers $r e^{i \varphi}$ one has $\varphi \in [0,2 \pi)$). In eq. \eqref{regul2} $l$ stands for additional regularization parameter chosen so that
\be
\label{el}
l =0,...,q-1.
\ee
We accept this temporarily without explanation; this will be clear in a moment. Note that for a given pair $(p,q)$ we have maximally $q-1$  cut-offs $\Lambda_{q,l}(N)$. The sense of introducing the forms \eqref{regul1}-\eqref{regul2} consists in the fact the norm \eqref{creg1} behaves well in the limit $N\rightarrow \infty$ due to the following periodicity:
\begin{align}
\nonumber
L_\lambda [\Lambda_{q,l}(N)] &= 1 + 2 \sin^{-1} \left(\frac{p}{q}\pi \right) \sin\left( l \frac{p}{q}\pi+N p \pi \right)  \cos\left( (l+1) \frac{p}{q}\pi+N p \pi \right) = 
\\[1ex]
& = 1 + 2 \sin^{-1} \left(\frac{p}{q}\pi \right) \sin\left( \frac{l \, p}{q}\pi \right) \cos\left( (l+1) \frac{p}{q}\pi \right).
\label{nlim}
\end{align}
By choosing the angle $\varphi$ to be rational number $p \pi/q$ and taking the cut-off $\Lambda_{q,k}(N)= q N +l$ we simply guaranty $L_\lambda [\Lambda_{q,l}(N)]$ does not depend on $N$ and, in consequence, behaves well in the limit $N \rightarrow \infty$. It is now clear what is the origin of the constraint \eqref{el}: any $\Lambda_{q,l}$ corresponding to $l \geq q$ is equivalent to $\Lambda_{q,l}$ corresponding to $l \leq q-1$. For instance, according to eq. \eqref{el} for $q=3$, one gets only three functions: $\Lambda_{3,0}=3 N$, $\Lambda_{3,1}=3 N + 1$, $\Lambda_{3,0}=3 N+2$. Now, suppose we ignore the upper bound, allowing more of them. For the fourth function one gets the cut-off $\Lambda_{3,4}=3 N+3 = 3 (N+1) = 3 \tilde{N}$, where $\tilde{N} := N+1$. However, this is trivially equivalent to $\Lambda_{3,0}$.

Having said that, we rewrite the initial form of eigenvectors \eqref{imgen} as
\be
\label{imgen2}
| e_{\lambda}^{(l)} \rangle  = \lim_{N \rightarrow \infty} E_{\lambda}^{\Lambda_{q,l} (N)} | \phi \rangle,
\ee
where $\lambda = r e^{i p \pi /q}$. It is easily to check that the vectors \eqref{imvec} are the special case of \eqref{imgen2}, corresponding to $r=1$, $(p,q) = (1,2)$ and $l=0,1$.

So far we restricted to normalization conditions and said nothing about the product of two different vectors. Using the rule \eqref{cel}, one finds
\begin{align}
\nonumber
\langle e_{\lambda_i}^{(l_i)} | e_{\lambda_j}^{(l_j)} \rangle = 1 &+ \sin^{-1} \left( \frac{\varphi_i+\varphi_j}{2}  \right) \lim_{N_i \rightarrow \infty} \sin \left(  \frac{\varphi_i+\varphi_j}{2} \Lambda_i(N_i)  \right)  \cos \left(  \frac{\varphi_i+\varphi_j}{2} (\Lambda_i(N_i)+1)  \right) +
\\[1ex]
&+ \sin^{-1} \left( \frac{\varphi_i+\varphi_j}{2}  \right) \lim_{N_j \rightarrow \infty} \sin \left(  \frac{\varphi_i+\varphi_j}{2} \Lambda_j(N_j)  \right)  \cos \left(  \frac{\varphi_i+\varphi_j}{2} (\Lambda_j(N_j)+1)  \right),
\label{cdprod}
\end{align}
where $\varphi_{i,j} = p_{i,j} \pi /q_{i,j}$, $\Lambda_{i,j} = q_{i,j} N_{i,j} + l_{i,j}$. The product \eqref{cdprod} was obtained under the assumption that $r_i = r_j$ (here $\lambda_{i} = r_{i} e^{i p_{i} \pi /q_{i}}$). If $r_i \neq r_j$ then $|\langle e_{\lambda_i}^{(l_i)} | e_{\lambda_j}^{(l_j)} \rangle| = \infty$. This can be easily read off from the original form of the product \eqref{Eij}. In what follows, eigenvectors labeled by eigenvalues of different $r$ cannot belong to the same (extended) Hilbert space because their product diverges. As we shall see in a moment, there are even stronger constrains on the eigenvalues. 

Before we go any further, it is worth mentioning that eigenvectors from different eigenvalues $\lambda_i$, $\lambda_j$ are orthogonal unless the eigenvalues are mutually conjugated, i.e. $\lambda_i = \lambda_j^*$. Indeed,
\begin{align}
\label{econ1}
&\langle e_{\lambda_j}| e|e_{\lambda_i}\rangle = \lambda_i \langle e_{\lambda_j}| e_{\lambda_i}\rangle,
\\[1ex]
\label{econ2}
&\langle e_{\lambda_j}| e|e_{\lambda_i}\rangle = (\langle e_{\lambda_i}| e^{\dagger}|e_{\lambda_j}\rangle)^* = \lambda_j^* \langle e_{\lambda_j}| e_{\lambda_i}\rangle.
\end{align}
Combining eqs. \eqref{econ1} and \eqref{econ2}, one concludes that either $\lambda_i = \lambda_j^*$ or $ \langle e_{\lambda_j}| e_{\lambda_i}\rangle = 0$. This is a direct analogue of the fact that for a Hermitian operator eigenvectors from different eigenvalues are orthogonal. Here we should keep in mind that allowing general complex eigenvectors we {\it do not} expect that $e$ is Hermitian in the corresponding space. Still, there is a mentioned consistency constraint. 

Looking closer at the product \eqref{cdprod} one concludes that $i)$ the product is, in general, not well-defined in the limit $N_m \rightarrow \infty$, $ii)$ vectors corresponding to two different eigenvalues $\lambda_i \neq  \lambda_j^*$ are not orthogonal even if the product is well-defined. The former can be circumvented modifying eq. \eqref{regul1} by making $p_m$ two times bigger, however, the latter would be still problematic. The same holds for eigenvectors corresponding to  $\lambda_i = \lambda_j^*$: the product is ill-defined. For instance, for $\lambda_i = r e^{i \varphi}$ ($r>0$) and  $\lambda_j = e^{-i \varphi}$ the product diverges. Hence, the only way to construct the space of complex normalized eigenvectors is restricting to a fixed eigenvalue. This leads to a superselection rule, forbidding superpositions of states corresponding to different complex eigenvalues. Note that this rule is stronger than the corresponding rule for the restricted Hilbert space \eqref{eps}. In the latter case we allowed superpositions of states corresponding to the same absolute value $|\lambda|$. Now we require the eigenvalues must be equal. In consequence, all the superselection sectors are built out of degenerated states.

Letting in \eqref{cdprod} $\lambda_i = \lambda_j = \lambda$, one finds
\be
\langle e_{\lambda}^{(l_i)} | e_{\lambda}^{(l_j)} \rangle = \sin^{-1}\left( \frac{p \pi}{q} \right) \cos \left( \frac{(l_i-l_j) p \pi}{q} \right) \sin \left( \frac{(1+l_i+l_j) p \pi}{q} \right).
\label{iprod}
\ee
Now the product \eqref{iprod} is well-defined and, in particular, does not depend on $N_{i}$ (which eventually goes to infinity). Notice that eigenvectors from a fixed eigenvalue $\lambda_1$ do not belong to the superselection sector corresponding to the eigenvalue $\lambda_2$, etc. The last is in parallel analogy to construction of regularized eigenvectors corresponding to real eigenvalues. The main difference is that each space corresponds to a single eigenvalue. 

Constructing the spaces of regularized eigenvectors we need to fix the eigenvalue $\lambda = r e^{i p \pi/q}$ first, and then find the corresponding eigenvectors. Since they are degenerated, any linear combination corresponding to the same eigenvalue $\lambda = r e^{i p \pi/q}$ is also a well-defined eigenvector (with the same eigenvalue). The number of positive, negative and zero norm eigenvectors is given by the signature of the quadratic form:
\be
\label{quad}
Q_{k n} = \langle e_{\lambda}^{(l_k)} | e_{\lambda}^{(l_n)} \rangle.
\ee
It turns out, the form above has a signature $\{-1,1,\vec{0}_{q-2} \}$, where $\vec{0}_{q-2}$ stands for a $q-2$ dimensional vector composed of zeros, $\vec{0}_{q-2} = \{ 0_{(1)},...,0_{(q-2)}\}$. If $q\leq 2$, then there is no zero eigenvectors. For instance, for $\lambda = e^{i \pi /4}$, one identifies $p=1$, $q=4$. Recalling eqs. \eqref{regul2} and \eqref{el}, one finds the following cut-offs
\be
\Lambda_{4,0} = 4 N, \, \Lambda_{4,1} = 4 N +1, \, \Lambda_{4,2} = 4 N +2, \, \Lambda_{4,3} = 4 N +3.
\ee
The corresponding regularized eigenvectors are given by eq. \eqref{imgen2}. This leads to
\be
\label{mat}
(Q_{k n})=\begin{bmatrix}
1& 1& 0& 0 \\
1& 1& 0& 0 \\
0& 0& -1& -1 \\
0& 0& -1& -1
\end{bmatrix}.
\ee
The matrix \eqref{mat} has the following eigenvalues $\{-2,2,0,0 \}$. Therefore, linear combinations of degenerated eigenvectors $| e_{\lambda}^{(l=0,...,3)} \rangle$, where $\lambda = e^{i \pi/4}$ result in one positive, one negative and two zero norm vectors (the signature is $(-1,1,0,0)$). In general, depending on the value of complex $\lambda$, the number of zero norm eigenvectors could be bigger or smaller, however, we get always two normalizable eigenvectors. 

Because different spaces of regularized eigenvectors differs only by the number of zero norm  vectors, they are all equivalent (in the sense that zero norm eigenvectors do not contribute to the  scalar product). Again, this is in parallel analogy to the case of real eigenvalues. As we recall, in this case the spaces are labeled by positive parameter $\lambda_0>0$, determining the pair of opposite eigenvalues $\pm \lambda_0$. In the complex case the states are degenerated and both positive and negative norm eigenvectors are present.

\end{document}